\documentclass[12pt]{article}  
\usepackage{cite}
\usepackage{epsfig}
\usepackage{graphicx}
\usepackage{amsmath}
\usepackage{amssymb}
\usepackage{ulem}
\usepackage{mdwlist} 
\usepackage{color}  

\usepackage{a41}
\usepackage{color}
\usepackage[rflt]{floatflt}
\usepackage{float}
\usepackage{slashed}

\usepackage{array}

\usepackage[english]{babel}
\usepackage[latin1]{inputenc}
\usepackage[T1]{fontenc}
\usepackage{ae}

\usepackage{url}

\usepackage{amsmath, amsthm, amssymb}

\newtheorem{thm}{Theorem}[section]

\newtheorem{definition}[thm]{Definition}

 \newcommand{\GeV}{\mathrm{GeV}}

 \newcommand{\MS}{\overline{\sf MS}}

 \newcommand{\Athathat}{\hat{\hspace*{0mm}\hat{\tilde{A}}}}

\newcommand{\NN}{\nonumber}

\newcommand{\ep}{\varepsilon}

\usepackage{rotating}

\usepackage{graphicx}

\newcounter{mmacnt}
\def\restartmma{\setcounter{mmacnt}{0}}
\restartmma \catcode`|=\active
\def|#1|{\mathrm{#1}}
\catcode`|=12
\newenvironment{mma}{
 \par\smallskip
 \catcode`|=\active
 \parskip=0pt\parindent=0pt 
 \small
 \def\In##1\\{%
\def\linebreak{\hfill\break\null\qquad}%
\refstepcounter{mmacnt}
\hangindent=2.5em\hangafter=0
\leavevmode
\llap{\tiny\sffamily n[\arabic{mmacnt}]:=\kern.5em}%
\mathversion{bold}\footnotesize$\displaystyle##1$\normalsize
\mathversion{normal}\par
 }%
 \def\Print##1\\{%
\def\linebreak{\hfill\break}%
\hangindent=2.5em\hangafter=0
\leavevmode ##1\par}%
 \def\Out##1\\{%
\def\linebreak{$\hfill\break\null\hfill$}%
\kern\abovedisplayskip\par
\hangindent=2.5em\hangafter=0
\leavevmode
\llap{\tiny\sffamily Out[\arabic{mmacnt}]=\kern.5em}
\footnotesize$\displaystyle##1$\normalsize\hfill\null\par
\kern\belowdisplayskip
 }%
 \def\Warning##1##2\\{%
\def\linebreak{\hfill\break}%
\hangindent=2.5em\hangafter=0
\leavevmode
{\scriptsize##1 : ##2}\par}%
}{%
 \par\smallskip
}

\usepackage{color}

\newenvironment{fshaded}{%
\MakeFramed {\FrameRestore}
}%
{\endMakeFramed}

\allowdisplaybreaks[4]

\begin{document}
\setlength{\baselineskip}{0.515cm}
\sloppy
\thispagestyle{empty}
\begin{flushleft}
DESY 19-163 
\\
DO-TH 19/19\\
TTP19-038 \\
SAGEX-19-23-E\\
November  2019\\
\end{flushleft}

\mbox{}
\vspace*{\fill}
\begin{center}

{\LARGE\bf The three-loop polarized pure singlet operator} 

\vspace*{3mm} 
{\LARGE\bf  matrix element with two different masses}

\vspace{3cm}
\large
J.~Ablinger$^a$, 
J.~Bl\"umlein$^b$, 
A.~De Freitas$^b$, 
M.~Saragnese$^b$, \\
C.~Schneider$^a$,   and  K.~Sch\"onwald$^{b,c}$ 

\vspace{1.cm}
\normalsize
{\it $^a$~Research Institute for Symbolic Computation (RISC),\\
  Johannes Kepler University, Altenbergerstra{\ss}e 69,
  A--4040, Linz, Austria}\\

\vspace*{3mm}
{\it  $^b$ Deutsches Elektronen--Synchrotron, DESY,}\\
{\it  Platanenallee 6, D-15738 Zeuthen, Germany}
\\

\vspace*{3mm}
{\it  $^c$ Institut f\"ur Theoretische Teilchenphysik, Campus S\"ud,\\
       Karlsruher Institut f\"ur Technologie (KIT), D-76128, Germany}
\\

\end{center}
\normalsize
\vspace{\fill}
\begin{abstract}
\noindent
We present the two-mass QCD contributions to the polarized pure singlet operator matrix element at three loop order 
in $x$-space. These terms are relevant for calculating the polarized structure function $g_1(x,Q^2)$ at $O(\alpha_s^3)$
as well as for the matching relations in the variable flavor number scheme and the polarized heavy quark distribution 
functions at the same order. The result for the operator matrix element is given in terms of generalized iterated 
integrals. These integrals depend on the mass ratio through the main argument, and the alphabet includes square--root 
valued letters. 
\end{abstract}

\vspace*{\fill}
\noindent
\numberwithin{equation}{section}
\newpage 
\section{Introduction}
\label{sec:1}

\vspace*{1mm}
\noindent
Massive operator matrix elements (OMEs) are essential building blocks for the massive Wilson coefficients
in deep--inelastic scattering in the limit $Q^2 \gg m^2$, and they are the transition matrix elements in the variable
flavor number scheme (VFNS) \cite{Buza:1995ie,Buza:1996wv}. Here $Q^2$ denotes the virtuality of the deep-inelastic 
process and $m$ is the heavy quark mass. From 2--loop order onward these matrix elements $A_{ij}$ receive two--mass 
corrections. In the unpolarized case the two--mass corrections have been calculated for all OMEs to three--loop order 
in Refs.~\cite{MOM,Ablinger:2017err,Ablinger:2017xml,Ablinger:2018brx,KS:THES}. For the OME $A_{Qg}^{(3),\rm tm}$ a large set 
of Mellin moments for even values of the Mellin variable $N \in \mathbb{N}$ has been derived, expanding in the mass ratio
\begin{eqnarray}
\eta = \frac{m_c^2}{m_b^2},
\end{eqnarray}
to a finite power, where $m_{c(b)}$ denote the charm and bottom quark mass, respectively.

In the polarized case, the flavor non--singlet three loop OME $A_{qq,Q}^{(3),\rm NS, tm}$ \cite{Ablinger:2017err} and 
$A_{gq,Q}^{(3),\rm tm}$ \cite{KS:THES} have been calculated. In the present paper we compute the two mass contributions 
to the pure singlet massive OME $A_{Qq}^{(3), \rm PS,  tm}$. Like in the unpolarized case, the calculation cannot be 
performed in $N$ space, transforming to momentum fraction $x$ space later, because the associated recurrences
do not factorize to first order. This, however, is the case for the corresponding differential equations in $x$
space. In the result we obtain iterative integrals, partly with limited support in $x \in [0,1]$. This has also been 
observed in the single mass pure singlet case \cite{Ablinger:2014nga}.

Since we will use dimensional regularization in the calculation, a consistent description of the Dirac matrix
$\gamma_5$ is necessary. For this we use the Larin scheme \cite{Larin:1993tq}.
The polarized massive OME $A_{Qq}^{(3), \rm PS, \rm tm}$ contributes to the polarized three-loop massive Wilson 
coefficient $H_{Qq}^{(3)}(z,Q^2)$ and is one of the contributions of the two--mass variable flavor number scheme
\cite{Ablinger:2017err,Blumlein:2018jfm} in the polarized case, describing the respective transitions of the polarized parton 
densities in the case the heavy quarks become light. They contribute in particular also to the charm- and bottom quark 
distributions.

The paper is organized as follows. In Section~\ref{sec:2} we present the renormalized pure-singlet OME in the 2-mass 
case. Details of the calculation are given in Section~\ref{sec:3}. In many steps of the calculation we follow the 
computation performed in Ref.~\cite{Ablinger:2017xml} in the unpolarized case, to be able to use 
the function space and the integral
relations which have been developed there. In Section~\ref{sec:4} the result of the 
calculation is presented and numerical results are given in Section~\ref{sec:5}. Section~\ref{sec:6} contains the 
conclusions. In the appendix we provide complete analytic expressions for a number of Mellin moments $N \in 
\mathbb{N}$.

\section{The renormalized 2-mass pure singlet OME}
\label{sec:2}

\vspace*{1mm}
\noindent
The generic pole structure of the polarized {\sf PS} three--loop two--mass contribution to the massive OME 
is given by \cite{Ablinger:2017err}
\begin{eqnarray}
\Athathat_{Qq}^{(3),\rm{PS, tm}} &=&
\frac{8}{3 \ep^3} \gamma_{gq}^{(0)} \hat{\gamma}_{qg}^{(0)} \beta_{0,Q}
+\frac{1}{\ep^2}\biggl[
2 \gamma_{gq}^{(0)} \hat{\gamma}_{qg}^{(0)} \beta_{0,Q} \left(L_1+L_2\right)
+\frac{1}{6} \hat{\gamma}_{qg}^{(0)} \hat{\gamma}_{gq}^{(1)}
-\frac{4}{3} \beta_{0,Q} \hat{\gamma}_{qq}^{\rm{PS},(1)}
\biggr] 
\NN\\&&
+\frac{1}{\ep}
\biggl[
\gamma_{gq}^{(0)} \hat{\gamma}_{qg}^{(0)} \beta_{0,Q} \left(L_1^2+L_1 L_2+L_2^2\right)
+\biggl\{\frac{1}{8} \hat{\gamma}_{qg}^{(0)} \hat{\gamma}_{gq}^{(1)}
- \beta_{0,Q} \hat{\gamma}_{qq}^{\rm{PS},(1)}
\biggr\} \left(L_2+L_1\right)
\NN\\&&
+\frac{1}{3} \hat{\tilde{\gamma}}_{qq}^{(2),\rm{PS}}
-8 a_{Qq}^{(2),\rm{PS}} \beta_{0,Q}
+ \hat{\gamma}_{qg}^{(0)} a_{gq}^{(2)}
\biggr] 
+\tilde{a}_{Qq}^{(3),\rm{PS}}\left(m_1^2,m_2^2,\mu^2\right)
                   \label{Ahhhqq3PSQ},
\end{eqnarray}
where we used the short hand notation
\begin{eqnarray}
\hat{\gamma}_{ij} &=& \gamma_{ij}(N_F+2) - \gamma_{ij}(N_F),
\\
\hat{\tilde{\gamma}}_{ij} &=& \frac{\gamma_{ij}(N_F+2)}{N_F+2}  - \frac{\gamma_{ij}(N_F)}{N_F}.
\end{eqnarray}
The tilde in $\Athathat_{Qq}^{(3),\rm{PS}}$ indicates that we are considering only the genuine two-mass contributions, and the double hat is used
to denote a completely unrenormalized OME. Here the $\gamma_{ij}^{(l)}$'s are anomalous dimensions at $l+1$ loops
\cite{Mertig:1995ny,SP_PS1,Moch:2014sna,Behring:2019tus}, $\beta_{0,Q}=-\frac{4}{3} T_F$, and
\begin{equation}
L_1 = \ln\left(\frac{m_1^2}{\mu^2}\right), \quad L_2 = \ln\left(\frac{m_2^2}{\mu^2}\right),
\label{L1L2}
\end{equation}
where $m_1$ and $m_2$ are the masses of the heavy quarks, and $\mu$ is the renormalization scale. 
Our goal is to compute the $O(\ep^0)$ term $\tilde{a}_{Qq}^{(3),\rm{PS}}\left(m_1^2,m_2^2,\mu^2\right)$.

We renormalize the heavy masses on-shell and  the coupling constant in the $\overline{\sf MS}$ scheme. 
The polarized OME in the Larin scheme is given by
\begin{eqnarray}
\tilde{A}_{Qq}^{(3), \MS, \rm{PS, tm}} &=&
-\gamma_{gq}^{(0)} \hat{\gamma}_{qg}^{(0)} \beta_{0,Q} 
\left(\frac{1}{4} L_2^2 L_1+ \frac{1}{4} L_1^2 L_2+ \frac{1}{3} L_1^3+\frac{1}{3} L_2^3\right)
\NN\\&&
+\biggl\{
-\frac{1}{16} \hat{\gamma}_{qg}^{(0)} \hat{\gamma}_{gq}^{(1)}
+ \frac{1}{2}\beta_{0,Q} \hat{\gamma}_{qq}^{\rm{PS},(1)}
\biggr\}
\left(L_2^2+L_1^2\right)
\NN\\&&
+\biggl\{
4 a_{Qq}^{(2), \rm{PS}} \beta_{0,Q}
- \frac{1}{2} \hat{\gamma}_{qg}^{(0)} a_{gq}^{(2)}
-\frac{1}{4} \beta_{0,Q} \zeta_2 \gamma_{gq}^{(0)} \hat{\gamma}_{qg}^{(0)}
\biggr\} \left(L_1+L_2\right)
\NN\\&&
+8 \overline{a}_{Qq}^{(2), \rm{PS}} \beta_{0,Q}
- \hat{\gamma}_{qg}^{(0)} \overline{a}_{gq}^{(2)}
+\tilde{a}_{Qq}^{(3), \rm{PS}}\left(m_1^2,m_2^2,\mu^2\right)
\label{Aqq3PSQMSren}.
\end{eqnarray}
The transition relations for the renormalization of the heavy quarks in the $\overline{\sf MS}$-scheme
is given in \cite{Ablinger:2014nga}, Eq.~(5.100), but it only applies to the equal mass case since for the unequal mass
case the first contributions emerge at 3--loop order. In Eqs. (\ref{Ahhhqq3PSQ}) and (\ref{Aqq3PSQMSren}), 
$a_{Qq}^{(2), \rm{PS}}$ and $a_{gq}^{(2)}$ represent the $O(\ep^0)$ terms
of the two-loop OMEs $\hat{\hat{A}}_{Qq}^{(2), \rm{PS}}$ and  $\hat{\hat{A}}_{gq}^{(2)}$, respectively, 
while $\overline{a}_{Qq}^{(2), \rm{PS}}$ and $\overline{a}_{gq}^{(2)}$ represent the corresponding $O(\ep)$ terms,
cf.~Refs.~\cite{Buza:1996xr,POL19,Hasselhuhn:2013swa,Blumlein:2019zux,KS:THES}.
Here and in what follows, $\zeta_k,~k \in \mathbb{N}, k \geq 2$ denotes the Riemann $\zeta$-function at integer argument.
\section{Details of the calculation}
\label{sec:3}

\begin{figure}[ht]
\begin{center}
\begin{minipage}[c]{0.19\linewidth}
     \includegraphics[width=1\textwidth]{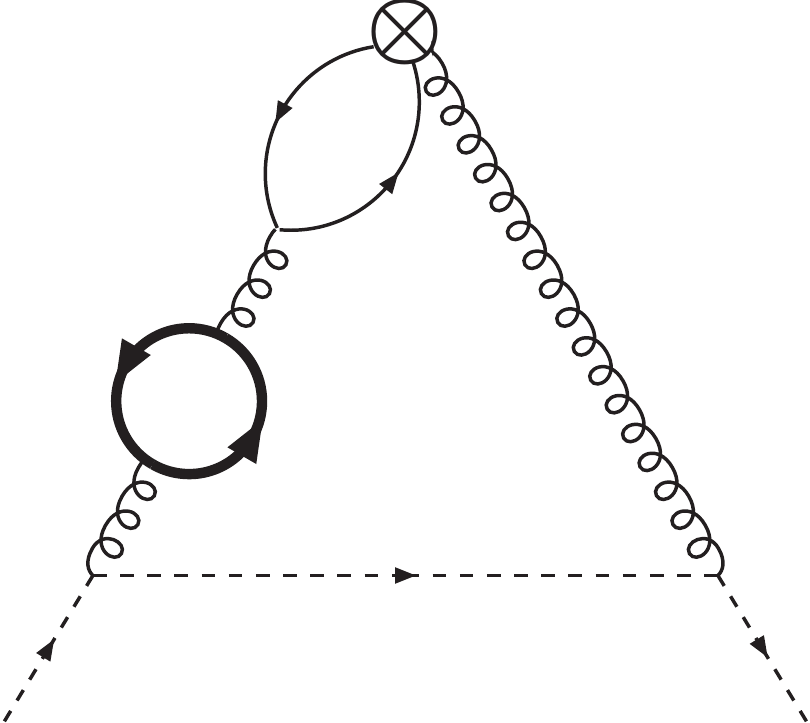}
\vspace*{-11mm}
\begin{center}
{\footnotesize (1)}
\end{center}
\end{minipage}
\hspace*{1mm}
\begin{minipage}[c]{0.19\linewidth}
     \includegraphics[width=1\textwidth]{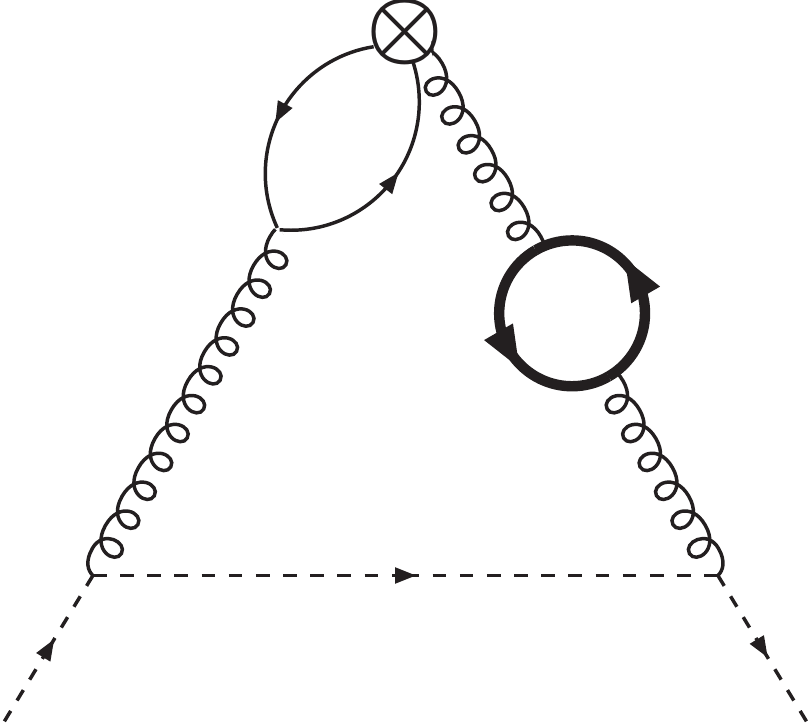}
\vspace*{-11mm}
\begin{center}
{\footnotesize (2)}
\end{center}
\end{minipage}
\hspace*{1mm}
\begin{minipage}[c]{0.19\linewidth}
     \includegraphics[width=1\textwidth]{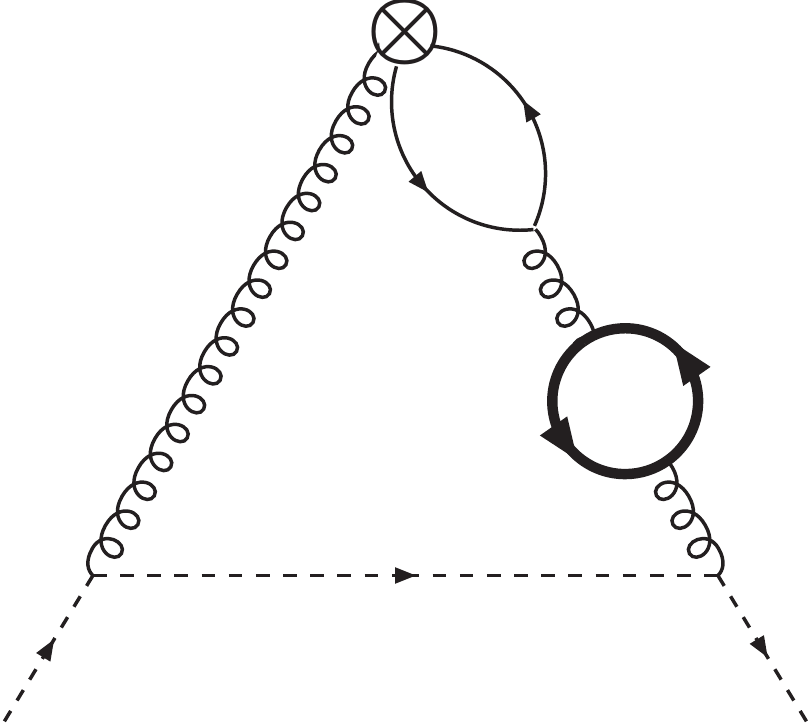}
\vspace*{-11mm}
\begin{center}
{\footnotesize (3)}
\end{center}
\end{minipage}
\hspace*{1mm}
\begin{minipage}[c]{0.19\linewidth}
     \includegraphics[width=1\textwidth]{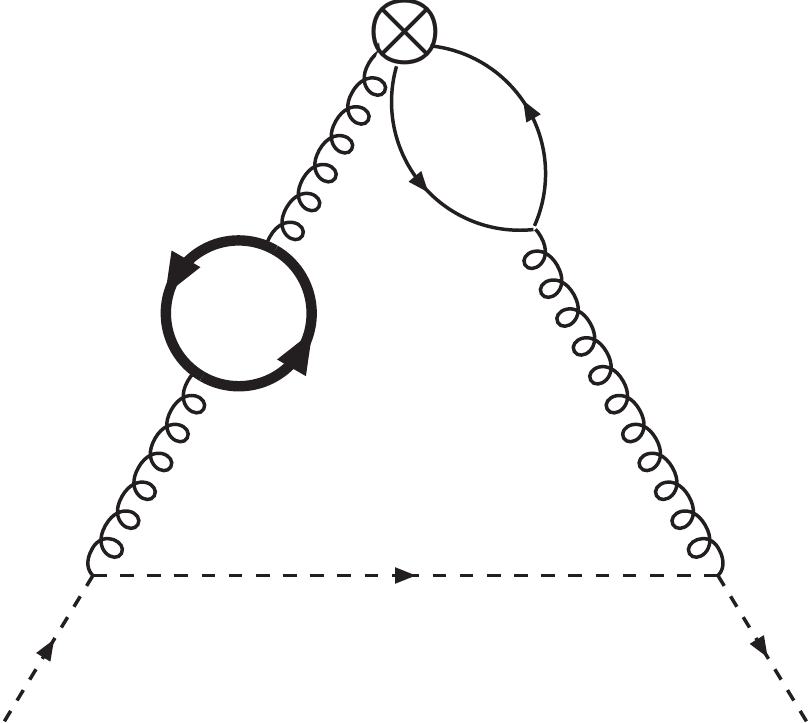}
\vspace*{-11mm}
\begin{center}
{\footnotesize (4)}
\end{center}
\end{minipage}

\vspace*{5mm}

\begin{minipage}[c]{0.19\linewidth}
     \includegraphics[width=1\textwidth]{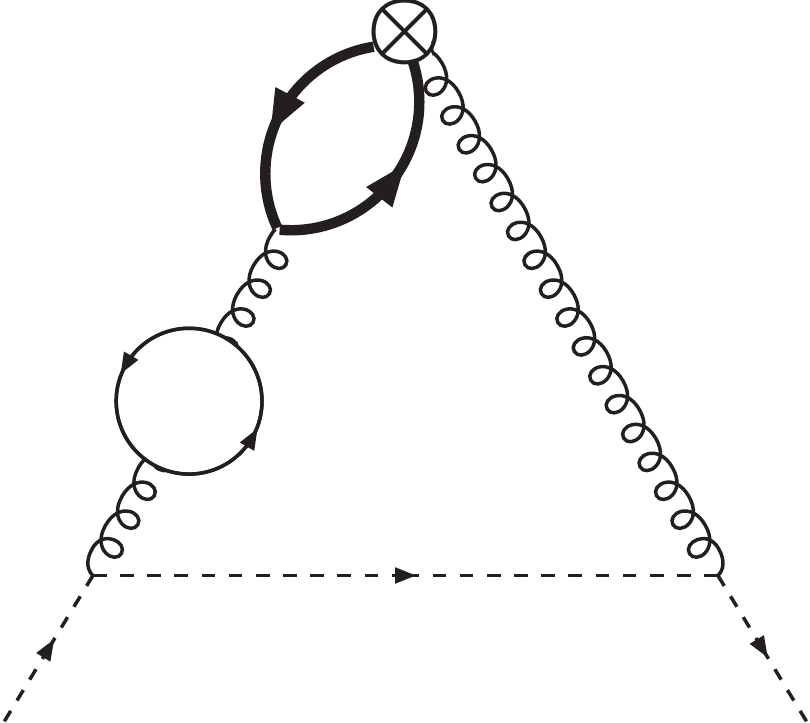}
\vspace*{-11mm}
\begin{center}
{\footnotesize (5)}
\end{center}
\end{minipage}
\hspace*{1mm}
\begin{minipage}[c]{0.19\linewidth}
     \includegraphics[width=1\textwidth]{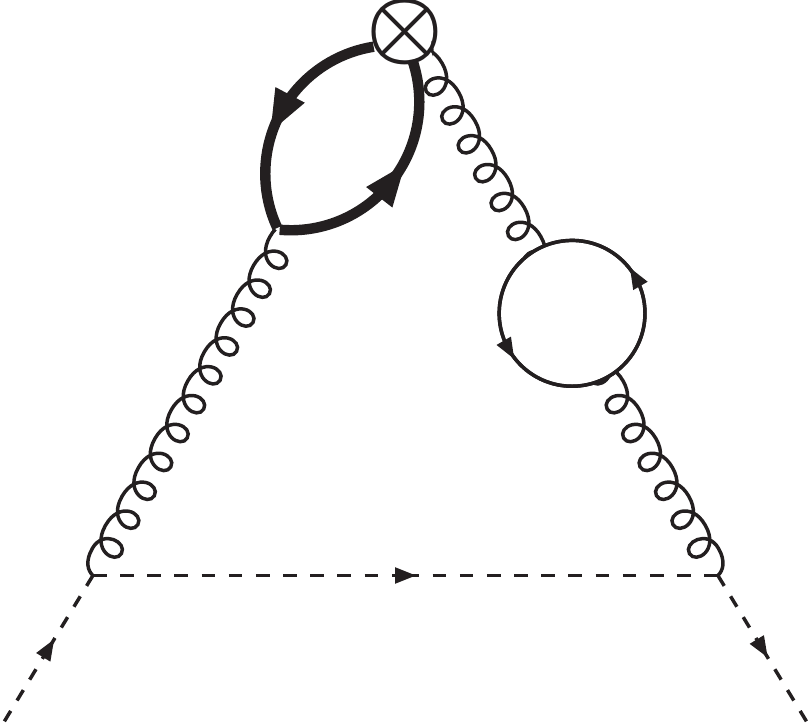}
\vspace*{-11mm}
\begin{center}
{\footnotesize (6)}
\end{center}
\end{minipage}
\hspace*{1mm}
\begin{minipage}[c]{0.19\linewidth}
     \includegraphics[width=1\textwidth]{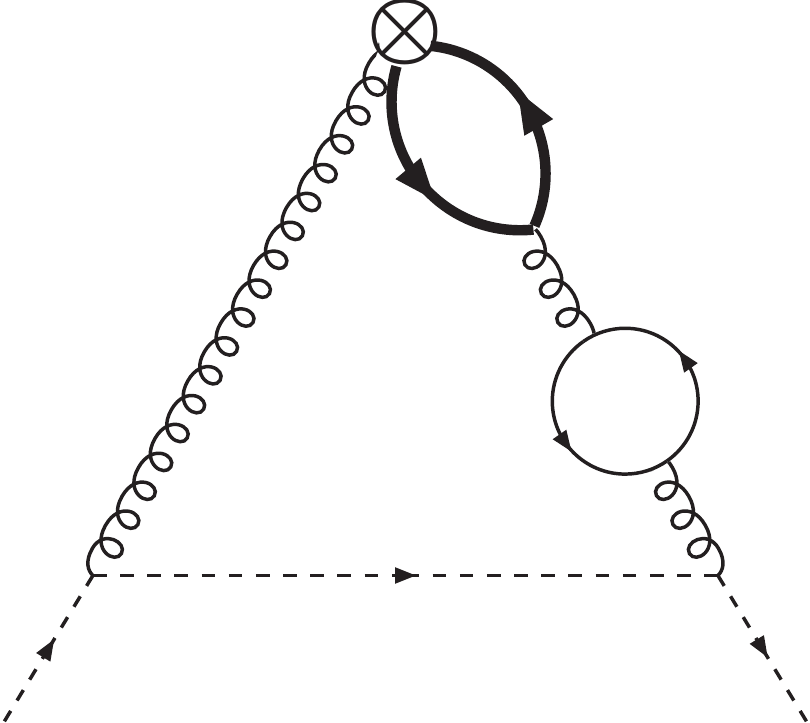}
\vspace*{-11mm}
\begin{center}
{\footnotesize (7)}
\end{center}
\end{minipage}
\hspace*{1mm}
\begin{minipage}[c]{0.19\linewidth}
     \includegraphics[width=1\textwidth]{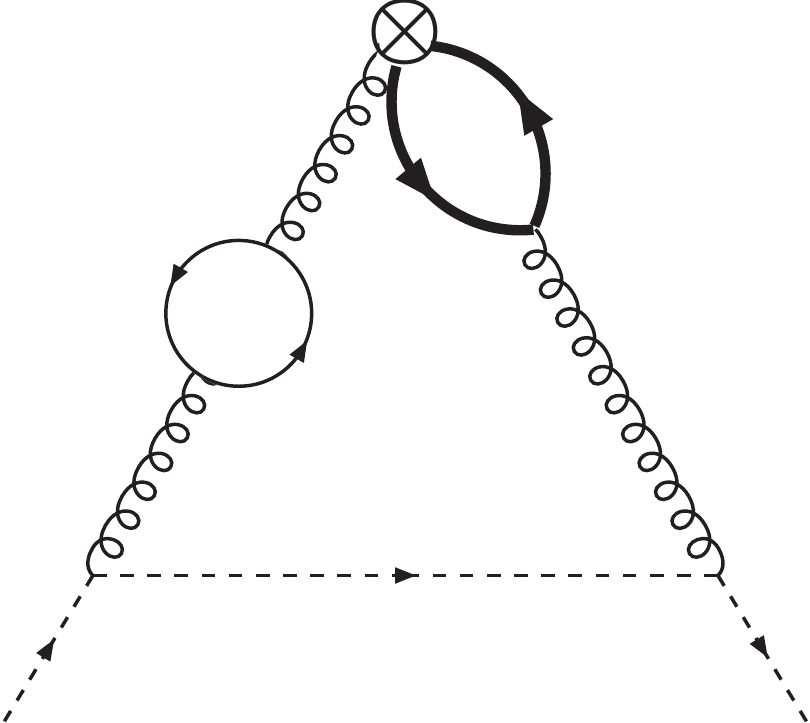}
\vspace*{-11mm}
\begin{center}
{\footnotesize (8)}
\end{center}
\end{minipage}

\vspace*{5mm}

\begin{minipage}[c]{0.19\linewidth}
     \includegraphics[width=1\textwidth]{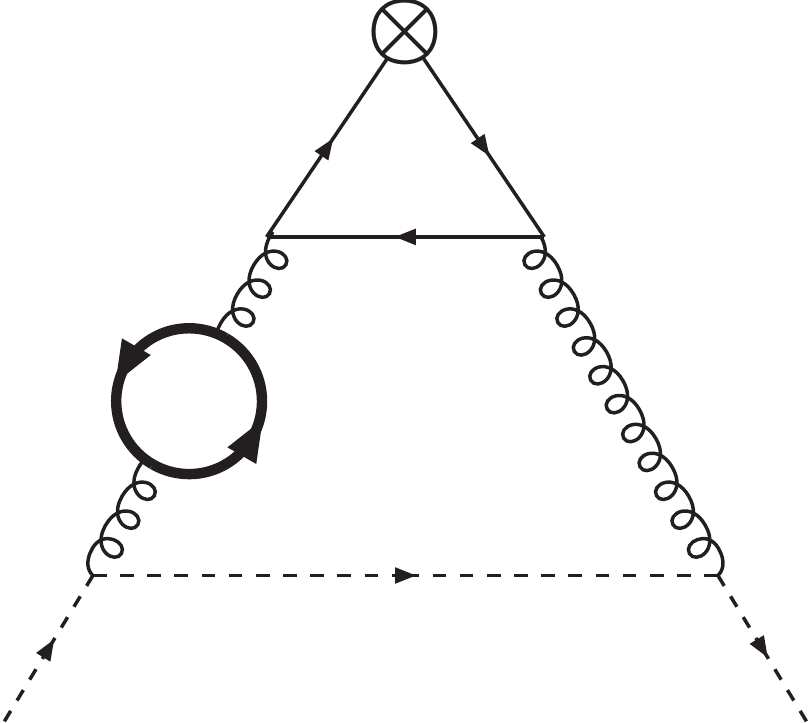}
\vspace*{-11mm}
\begin{center}
{\footnotesize (9)}
\end{center}
\end{minipage}
\hspace*{1mm}
\begin{minipage}[c]{0.19\linewidth}
     \includegraphics[width=1\textwidth]{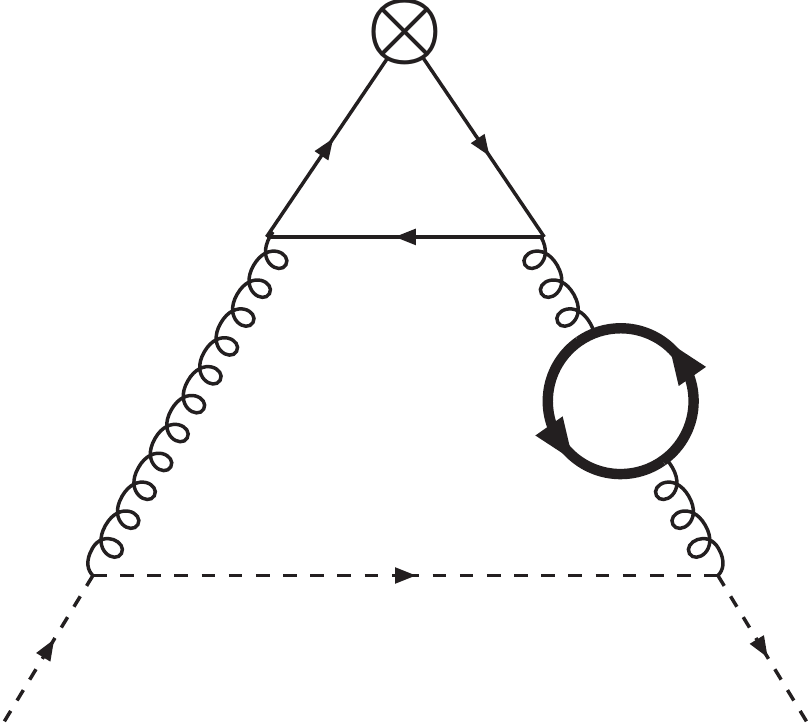}
\vspace*{-11mm}
\begin{center}
{\footnotesize (10)}
\end{center}
\end{minipage}
\hspace*{1mm}
\begin{minipage}[c]{0.19\linewidth}
     \includegraphics[width=1\textwidth]{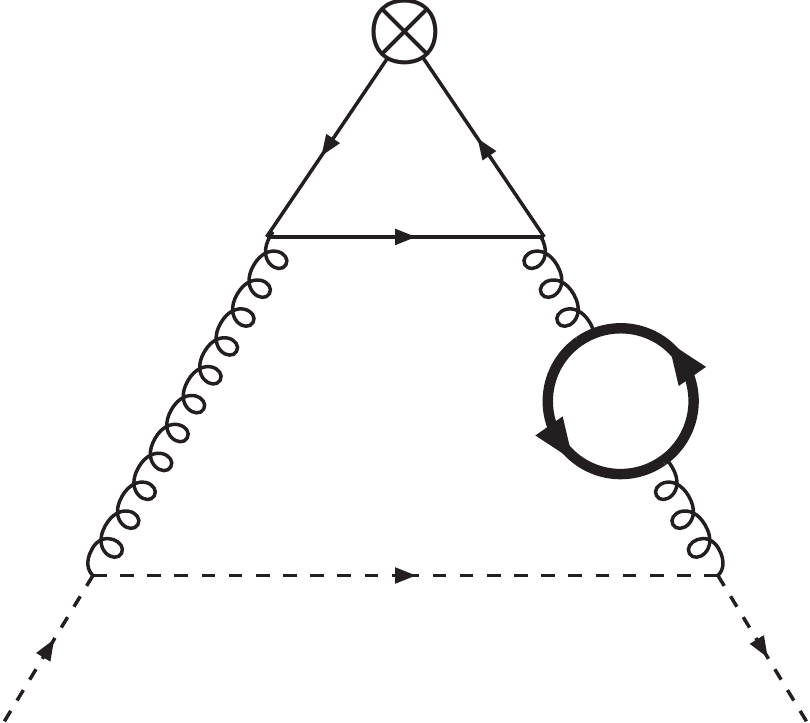}
\vspace*{-11mm}
\begin{center}
{\footnotesize (11)}
\end{center}
\end{minipage}
\hspace*{1mm}
\begin{minipage}[c]{0.19\linewidth}
     \includegraphics[width=1\textwidth]{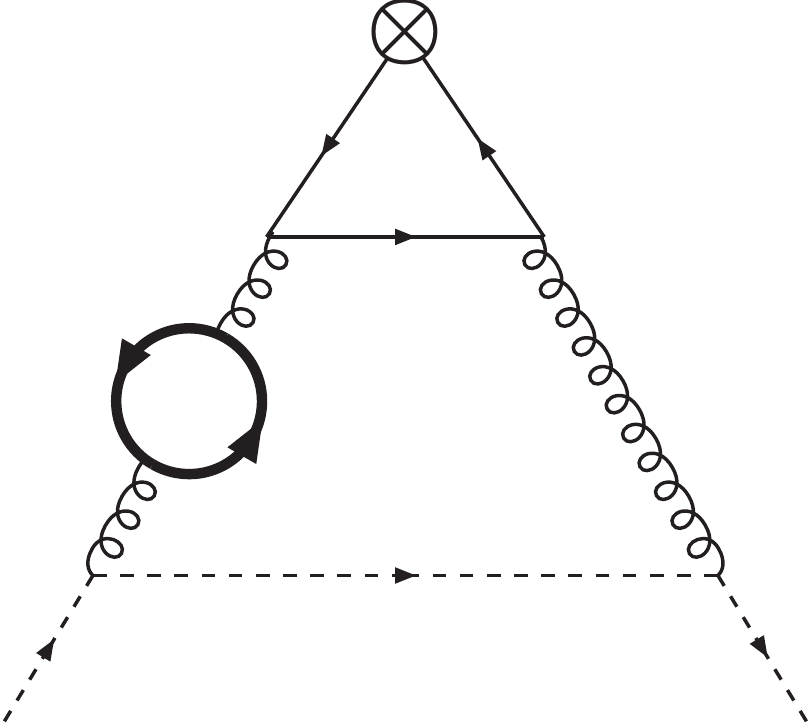}
\vspace*{-11mm}
\begin{center}
{\footnotesize (12)}
\end{center}
\end{minipage}

\vspace*{5mm}

\begin{minipage}[c]{0.19\linewidth}
     \includegraphics[width=1\textwidth]{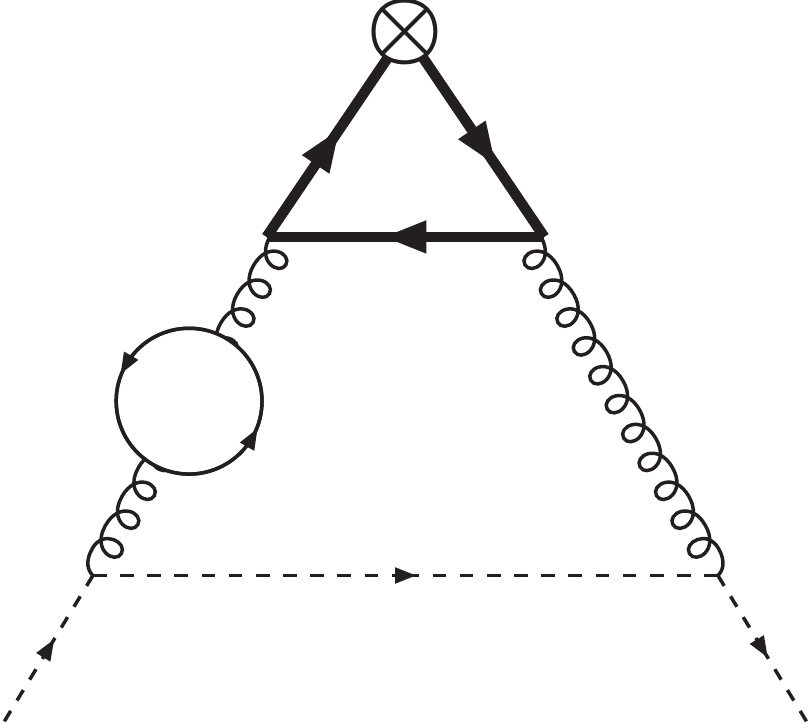}
\vspace*{-11mm}
\begin{center}
{\footnotesize (13)}
\end{center}
\end{minipage}
\hspace*{1mm}
\begin{minipage}[c]{0.19\linewidth}
     \includegraphics[width=1\textwidth]{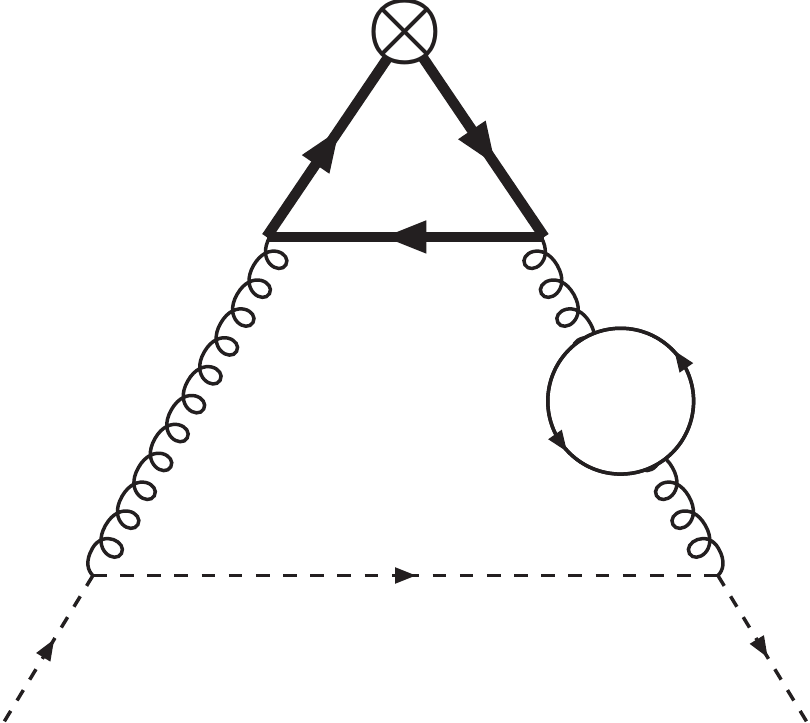}
\vspace*{-11mm}
\begin{center}
{\footnotesize (14)}
\end{center}
\end{minipage}
\hspace*{1mm}
\begin{minipage}[c]{0.19\linewidth}
     \includegraphics[width=1\textwidth]{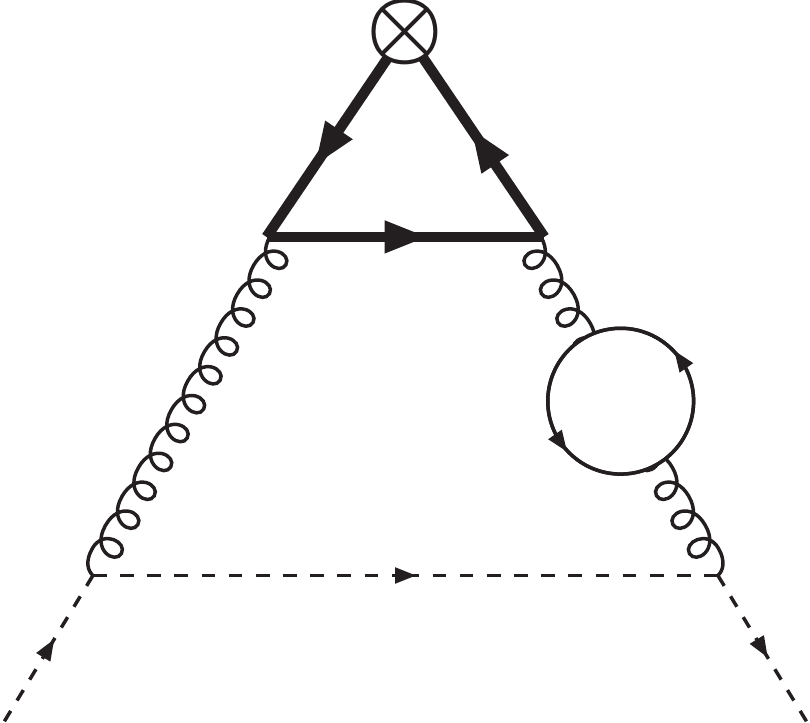}
\vspace*{-11mm}
\begin{center}
{\footnotesize (15)}
\end{center}
\end{minipage}
\hspace*{1mm}
\begin{minipage}[c]{0.19\linewidth}
     \includegraphics[width=1\textwidth]{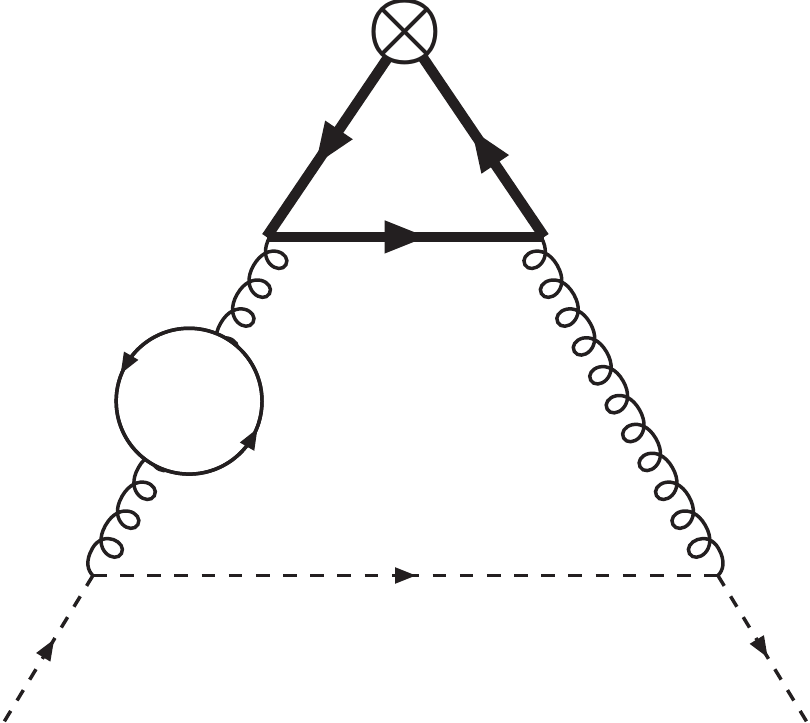}
\vspace*{-11mm}
\begin{center}
{\footnotesize (16)}
\end{center}
\end{minipage}
\caption{\sf \small The diagrams for the two-mass contributions to $\tilde{A}_{Qq}^{(3), \rm PS}$. The dashed arrow line represents the external
massless quarks, while the thick solid arrow line represents a quark of mass $m_1$, and the thin arrow line a quark of mass $m_2$. We assume $m_1 > m_2$.}
\label{diagrams}
\end{center}
\end{figure}

\vspace*{1mm}
\noindent
There are sixteen irreducible diagrams for $\tilde{A}_{Qq}^{(3), {\rm PS, tm}}$, which are shown in Figure~\ref{diagrams}. 
The unrenormalized operator matrix element is obtained by adding all the diagrams and applying the quarkonic projector $P_q$ to the 
corresponding Green function $\hat{G}^{ij}_Q$, cf.~\cite{Behring:2019tus},
\begin{eqnarray}
\label{eq:PqNEW}
P_q \hat{G}_l^{ij} = - \delta_{ij} \frac{i (\Delta.p)^{-N-1}}{4 N_c (D-2)(D-3)} \ep_{\mu \nu p \Delta} {\rm tr} \left[p
\hspace*{-2mm}
\slash
\gamma^\mu \gamma^\nu \hat{G}_l^{ij}\right],
\end{eqnarray}
where $p$ is the momentum of the on-shell external massless quark ($p^2=0$), $\Delta$ is a light-like $D$-vector, 
with $D = 4+ \varepsilon$, the dimension of space-time in which we work, $i$ and $j$ are the color indices of each external 
leg, and $N_c$ is the number of colors. 
Note that the projector~(\ref{eq:PqNEW}) is different from that in the unpolarized case \cite{Ablinger:2017xml}.
The diagrams, $D_1, \ldots, D_{16}$, are calculated directly within dimensional regularization.
The Dirac algebra is performed using {\tt FORM} \cite{FORM}.
Diagrams 1--8 turn out to vanish. Diagrams 9--12 and 13--16 can be mapped by symmetry relations to each other, 
respectively. These two classes are furthermore related by exchanging $\eta \Leftrightarrow 1/\eta$.

One therefore obtains
\begin{eqnarray}
A_{Qq}^{(3), {\rm PS, tm}}(N) &=&
2 \left[1 + (-1)^{N-1} \right] D_9(m_1,m_2,N) + 2 \left[1 + (-1)^{N-1} \right] D_9(m_2,m_1,N),
\label{eq:D9}
\end{eqnarray}
where $N$ is the Mellin variable appearing in the Feynman rules for the operator insertions, 
cf.~\cite{Mertig:1995ny,Behring:2019tus}. In the following we use the variable
\begin{equation}
\eta = \frac{m_2^2}{m_1^2},
\end{equation}
with $m_2 < m_1$, i.e.~$\eta < 1$, which we will assume in what follows.

While in other calculations one could derive the results working either in Mellin $N$ or $x$--space, cf.
e.g. \cite{Ablinger:2017err, Ablinger:2018brx}, this is not the case here, see also \cite{Ablinger:2017xml}.
We will, therefore, present our result only in $x$-space, which is anyway 
all we need in order to obtain the corresponding contribution to
the structure function $g_1(x,Q^2)$ for large values of $Q^2$, as well as the contribution to the variable 
flavor number scheme. In most of the applications one finally works in $x$-space. 
\begin{figure}[ht]
\begin{center}
\begin{minipage}[c]{0.21\linewidth}
     \includegraphics[width=1\textwidth]{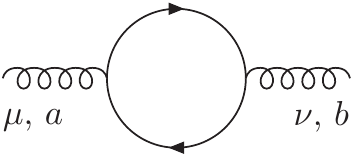}
\vspace*{-7mm}
\begin{center}
{\footnotesize ($a_1$)}
\end{center}
\end{minipage}
\hspace*{11mm}
\begin{minipage}[c]{0.21\linewidth}
     \includegraphics[width=1\textwidth]{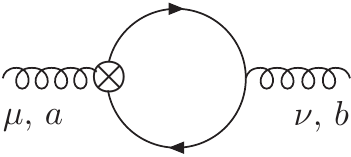}
\vspace*{-7mm}
\begin{center}
{\footnotesize ($b_1$)}
\end{center}
\end{minipage}
\hspace*{11mm}
\begin{minipage}[c]{0.21\linewidth}
     \includegraphics[width=1\textwidth]{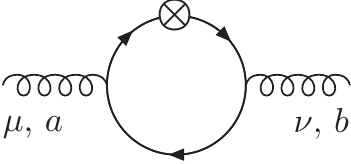}
\vspace*{-7mm}
\begin{center}
{\footnotesize ($b_2$)}
\end{center}
\end{minipage}
\caption{\sf \small Massive bubbles appearing in the Feynman diagrams shown in Figure \ref{diagrams}.}
\label{bubbles}
\end{center}
\end{figure}

All the diagrams contain a massive fermion loop with an operator insertion (Figures \ref{bubbles}$b_1$ and \ref{bubbles}$b_2$)
and a massive bubble without the operator (Figure \ref{bubbles}$a_1$).
The latter can be rendered effectively massless by using a Mellin--Barnes
integral~\cite{MB1a,MB1b,MB2,MB3,MB4}. One obtains
\begin{eqnarray}
I^{\mu\nu,ab}_{a_1}(k) &=&
-\frac{8iT_F g_s^2}{(4\pi)^{D/2}}\delta_{ab}(k^2g^{\mu\nu}-k^\mu k^\nu)\int_0^1 
dx\frac{\Gamma(2-D/2)(x(1-x))^{D/2-1}}{\left(-k^2+\frac{m^2}{x(1-x)}\right)^{2-D/2}},\\
I^{\mu\nu,ab}_{b_2}(k) &=&
\alpha_s T_F i e^{-\gamma_E \varepsilon/2}(k\cdot\Delta)^{N-1} (\mu^2)^{-\varepsilon/2}S_\varepsilon \epsilon^{\Delta k 
\mu 
\nu}\int_0^1 dx\, x^{N+D/2-1}(1-x)^{D/2-1}\nonumber\\
&&
\times\Bigg\{\left(-k^2+\frac{m^2}{x(1-x)}\right)^{-2+D/2}2\Gamma(2-D/2)\left[(D-6)x^{-2}+(D+2N)x^{-1}\right]\\\nonumber&&
+\left(-k^2+\frac{m^2}{x(1-x)}\right)^{-3+D/2}4\Gamma(3-D/2)(1-x)^{-1}\Big[m^2(x^{-3}+x^{-2})\nonumber\\
&&
+(-k^2)(1-x^{-1})\Big]\Bigg\},
\end{eqnarray}
where $\mu$ and $\nu$ are the respective Lorentz indices of the external legs, $a$ and $b$ are the color indices, 
$k$ is the external momentum, $m$ is the mass of the fermion, which can be either $m_1$ or 
$m_2$, $g_s = \sqrt{4 \pi \alpha_s}$ is the strong coupling constant, and $T_F=1/2$ in $SU(N_c)$, with $N_c$ the 
number of colors. The other color factors are $C_F = (N_c^2-1)/(2 N_c)$ and $C_A = N_c$.
The term $I_{b_1}^{\mu\nu, ab}(k)$ only appears in diagrams which vanish and is not displayed here.

For diagram 9 we obtain the representation
\begin{eqnarray}
D_{9}(m_{1},m_{2},N) & = & C_F T_F^2 \alpha_s^3 S_\varepsilon^3 \frac{16}{2+\varepsilon}\Bigg\{ 4(2-\varepsilon)J_{1}-8\eta 
J_{2}-8(N+3)J_{3}+8J_{4}+8\bigg(2+\frac{\varepsilon}{2}+N\bigg)
\nonumber\\ 
& &
\times J_{5}
-8J_{6}-(\varepsilon-2)^{2}J_{7}+2(2-\varepsilon)\eta J_{8}+2(2-\varepsilon)(3+N)J_{9}-2(2-\varepsilon)J_{10}
\nonumber\\
 &  & -2(2-\varepsilon)\bigg(2+\frac{\varepsilon}{2}+N\bigg)J_{11}+2(2-\varepsilon)J_{12}-8\eta J_{13}+2(2-\varepsilon)\eta 
J_{14}\Bigg\},
\end{eqnarray}
with
\begin{eqnarray}
J_{1} & = & \left(\frac{m_{1}^{2}}{\mu^{2}}\right)^{\frac{3}{2}\varepsilon}\frac{\Gamma(N)}{\Gamma\big(1+\frac{\varepsilon 
}{2}+N\big)}\int_{0}^{1}dx\,(1-x)^{\frac{\varepsilon}{2}}x^{-1+\frac{\varepsilon }{2}+N}B_{1}\left(\frac{\eta}{x(1-x)}\right),\\
J_{2} & = & \left(\frac{m_{1}^{2}}{\mu^{2}}\right)^{\frac{3}{2}\varepsilon}\frac{\Gamma(N)}{\Gamma\big(1+\frac{\varepsilon 
}{2}+N\big)}\int_{0}^{1}dx\,(1-x)^{\frac{\varepsilon}{2}}x^{-1+\frac{\varepsilon }{2}+N}B_{3}\left(\frac{\eta}{x(1-x)}\right),\\
J_{3} & = & \left(\frac{m_{1}^{2}}{\mu^{2}}\right)^{\frac{3}{2}\varepsilon}\frac{\Gamma(N)}{\Gamma\big(1+\frac{\varepsilon 
}{2}+N\big)}\int_{0}^{1}dx\,(1-x)^{\frac{\varepsilon}{2}}x^{\frac{\varepsilon }{2}+N}B_{1}\left(\frac{\eta}{x(1-x)}\right),\\
J_{4} & = & \left(\frac{m_{1}^{2}}{\mu^{2}}\right)^{\frac{3}{2}\varepsilon}\frac{\Gamma(N)}{\Gamma\big(1+\frac{\varepsilon 
}{2}+N\big)}\int_{0}^{1}dx\,(1-x)^{\frac{\varepsilon}{2}}x^{\frac{\varepsilon }{2}+N}B_{2}\left(\frac{\eta}{x(1-x)}\right),\\
J_{5} & = & \left(\frac{m_{1}^{2}}{\mu^{2}}\right)^{\frac{3}{2}\varepsilon}\frac{\Gamma(N)}{\Gamma\big(1+\frac{\varepsilon 
}{2}+N\big)}\int_{0}^{1}dx\,(1-x)^{\frac{\varepsilon}{2}}x^{1+\frac{\varepsilon }{2}+N}B_{1}\left(\frac{\eta}{x(1-x)}\right),\\
J_{6} & = & \left(\frac{m_{1}^{2}}{\mu^{2}}\right)^{\frac{3}{2}\varepsilon}\frac{\Gamma(N)}{\Gamma\big(1+\frac{\varepsilon 
}{2}+N\big)}\int_{0}^{1}dx\,(1-x)^{\frac{\varepsilon}{2}}x^{1+\frac{\varepsilon }{2}+N}B_{2}\left(\frac{\eta}{x(1-x)}\right),\\
J_{7} & = & \left(\frac{m_{1}^{2}}{\mu^{2}}\right)^{\frac{3}{2}\varepsilon}\frac{\Gamma(N+1)}{\Gamma\big(2+\frac{\varepsilon 
}{2}+N\big)}\int_{0}^{1}dx\,(1-x)^{\frac{\varepsilon}{2}}x^{-1+\frac{\varepsilon }{2}+N}B_{1}\left(\frac{\eta}{x(1-x)}\right),\\
J_{8} & = & \left(\frac{m_{1}^{2}}{\mu^{2}}\right)^{\frac{3}{2}\varepsilon}\frac{\Gamma(N+1)}{\Gamma\big(2+\frac{\varepsilon 
}{2}+N\big)}\int_{0}^{1}dx\,(1-x)^{\frac{\varepsilon}{2}}x^{-1+\frac{\varepsilon }{2}+N}B_{3}\left(\frac{\eta}{x(1-x)}\right),\\
J_{9} & = & \left(\frac{m_{1}^{2}}{\mu^{2}}\right)^{\frac{3}{2}\varepsilon}\frac{\Gamma(N+1)}{\Gamma\big(2+\frac{\varepsilon 
}{2}+N\big)}\int_{0}^{1}dx\,(1-x)^{\frac{\varepsilon}{2}}x^{\frac{\varepsilon }{2}+N}B_{1}\left(\frac{\eta}{x(1-x)}\right),\\
J_{10} & = & \left(\frac{m_{1}^{2}}{\mu^{2}}\right)^{\frac{3}{2}\varepsilon}\frac{\Gamma(N+1)}{\Gamma\big(2+\frac{\varepsilon 
}{2}+N\big)}\int_{0}^{1}dx\,(1-x)^{\frac{\varepsilon}{2}}x^{\frac{\varepsilon }{2}+N}B_{2}\left(\frac{\eta}{x(1-x)}\right),\\
J_{11} & = & \left(\frac{m_{1}^{2}}{\mu^{2}}\right)^{\frac{3}{2}\varepsilon}\frac{\Gamma(N+1)}{\Gamma\big(2+\frac{\varepsilon 
}{2}+N\big)}\int_{0}^{1}dx\,(1-x)^{\frac{\varepsilon}{2}}x^{1+\frac{\varepsilon }{2}+N}B_{1}\left(\frac{\eta}{x(1-x)}\right),\\
J_{12} & = & \left(\frac{m_{1}^{2}}{\mu^{2}}\right)^{\frac{3}{2}\varepsilon}\frac{\Gamma(N+1)}{\Gamma\big(2+\frac{\varepsilon 
}{2}+N\big)}\int_{0}^{1}dx\,(1-x)^{\frac{\varepsilon}{2}}x^{1+\frac{\varepsilon }{2}+N}B_{2}\left(\frac{\eta}{x(1-x)}\right),\\
J_{13} & = & \left(\frac{m_{1}^{2}}{\mu^{2}}\right)^{\frac{3}{2}\varepsilon}\frac{\Gamma(N)}{\Gamma\big(1+\frac{\varepsilon 
}{2}+N\big)}\int_{0}^{1}dx\,(1-x)^{\frac{\varepsilon}{2}}x^{-2+\frac{\varepsilon }{2}+N}B_{3}\left(\frac{\eta}{x(1-x)}\right),\\
J_{14} & = & \left(\frac{m_{1}^{2}}{\mu^{2}}\right)^{\frac{3}{2}\varepsilon}\frac{\Gamma(N+1)}{\Gamma\big(2+\frac{\varepsilon 
}{2}+N\big)}\int_{0}^{1}dx\,(1-x)^{\frac{\varepsilon}{2}}x^{-2+\frac{\varepsilon }{2}+N}B_{3}\left(\frac{\eta}{x(1-x)}\right).
\end{eqnarray}
The functions $B_i$ are given by
\begin{eqnarray}
B_1(\xi) &=& \frac{1}{2 \pi i} \int_{-i \infty}^{i \infty} d\sigma \, \xi^{\sigma} \,
\Gamma(-\sigma) \Gamma(-\sigma+\varepsilon) \Gamma\left(\sigma-\frac{3 \varepsilon}{2}\right) \Gamma\left(\sigma 
-\frac{\varepsilon}{2}\right)
\frac{\Gamma^2(\sigma+2-\varepsilon)}{\Gamma(2 \sigma+4-2 \varepsilon)},
\label{B1}
\\
B_2(\xi) &=& \frac{1}{2 \pi i} \int_{-i \infty}^{i \infty} d\sigma \, \xi^{\sigma} \,
\Gamma(-\sigma) \Gamma(-\sigma+\varepsilon) \Gamma\left(\sigma-\frac{3 \varepsilon}{2}\right) 
\Gamma\left(\sigma+1-\frac{\varepsilon}{2}\right)
\frac{\Gamma^2(\sigma+2-\varepsilon)}{\Gamma(2 \sigma+4-2 \varepsilon)},
\nonumber \\   
\label{B2}
\\
B_3(\xi) &=& \frac{1}{2 \pi i} \int_{-i \infty}^{i \infty} d\sigma \, \xi^{\sigma} \,
\Gamma(-\sigma) \Gamma(-\sigma-1+\varepsilon) \Gamma\left(\sigma+1-\frac{3 \varepsilon}{2}\right) 
\Gamma\left(\sigma+1-\frac{\varepsilon}{2}\right)
\nonumber \\ && \phantom{\frac{1}{2 \pi i} \int_{-i \infty}^{i \infty} d\sigma} \times
\frac{\Gamma^2(\sigma+3-\varepsilon)}{\Gamma(2 \sigma+6-2 \varepsilon)}.
\label{B3}
\end{eqnarray}
Expanding in the dimensional parameter $\varepsilon$ the pre-factors of the functions $J_i$ reduce to the following 
denominators 
\begin{equation}
\frac{1}{N+l}, \quad {\rm with} \quad l \in \{0,1\},
\end{equation}
after partial fractioning.
These factors have still to be absorbed under the integral, which can be achieved by applying the following relations
\begin{eqnarray}
\frac{1}{N+l} \int_a^b dx \, x^{N-1} f(x) &=&
\frac{b^{N+l}}{N+l} \int_a^b dy \frac{f(y)}{y^{l+1}} - \int_a^b dx \, x^{N+l-1} \int_a^x dy \frac{f(y)}{y^{l+1}}
\label{absorbN1}
\\ &=&
\frac{a^{N+l}}{N+l} \int_a^b dy \frac{f(y)}{y^{l+1}} + \int_a^b dx \, x^{N+l-1} \int_x^b dy \frac{f(y)}{y^{l+1}}.
\label{absorbN2}
\end{eqnarray}   

One has still to perform the contour integral in the functions $B_i$. To do this, the range in $x$ is split into the 
intervals
\begin{eqnarray}   
[0,\eta_-],~~~[\eta_-,\eta_+],~~~[\eta_+,1],~~~\text{with}~~\eta_\pm = \frac{1}{2}\left(1 \pm \sqrt{1-\eta}\right).
\end{eqnarray}   
For the second region the integral contour is closed to the right, and for the two other regions to the left. One obtains
then the functions $B_i$ for both regions in terms of infinite sum representations, cf.~\cite{Ablinger:2017xml}, over rational 
expressions and harmonic sums \cite{Vermaseren:1998uu,Blumlein:1998if}
\begin{eqnarray}
S_{b,\vec{a}}(N) = \sum_{k=1}^N \frac{({\rm sign}(b))^k}{k^{|b|}} S_{\vec{a}}(k),~~~S_\emptyset = 1,~~~b,a_i \in \mathbb{Z} 
\backslash \{0\}~.
\end{eqnarray}
In the expressions ratios of $\Gamma$-functions are related to special binomial coefficients, like
\begin{eqnarray}
\frac{\Gamma^2(k+1)}{\Gamma(2k+2)} = \frac{1}{2k} \frac{1}{\displaystyle \binom{2k}{k}}.
\end{eqnarray}
All of the above sums can be performed using the Mathematica packages {\tt Sigma}~\cite{SIG1,SIG2}, {\tt 
HarmonicSums}~\cite{HARMONICSUMS,Ablinger:2011te,Ablinger:2013cf},
{\tt EvaluateMultiSums} and {\tt SumProduction} \cite{EMSSP}.
We have performed numerical checks on these steps using the package {\tt MB} and {\tt MBresolve} \cite{MB,MBr}.

The results can be expressed in terms of the following generalized iterated integrals.
\begin{equation}
G\left(\left\{f_1(\tau),f_2(\tau),\cdots,f_n(\tau)\right\},z\right)
=\int_0^z  d\tau_1~f_1(\tau_1)  
G\left(\left\{f_2(\tau),\cdots,f_n(\tau)\right\},\tau_1\right),
\label{Gfunctions}
\end{equation}
with
\begin{equation}
 G\Biggl(\Biggl\{\underbrace{\frac{1}{\tau},\frac{1}{\tau},
  \cdots,\frac{1}{\tau}}_{\text{n times}}\Biggr\},z\Biggr)
\equiv
\frac{1}{n!} \ln^n(z)~.
\end{equation}
In principle, the letters in the alphabet of these iterated integrals, i.e., the functions $f_k(\tau)$, can be 
any function (or distribution), for which the iterated integral exists. 
In the particular case where the letters are restricted to $\frac{1}{\tau}$, $\frac{1}{1-\tau}$ and $\frac{1}{1+\tau}$, these
integrals correspond to the harmonic polylogarithms \cite{Remiddi:1999ew}, which are defined by
\begin{eqnarray}
H_{b,\vec{a}}(x) &=& \int_0^x dy f_b(y) H_{\vec{a}}(y),~~~H_\emptyset = 1,~a_i, b~\in~\{0,1,-1\}~,
\end{eqnarray}
with
\begin{eqnarray}
f_0(x) = \frac{1}{x},~~~f_1(x) = \frac{1}{1-x},~~~f_{-1}(x) = \frac{1}{1+x}~,
\end{eqnarray}
and
\begin{equation}
H_{\underbrace{\text{\scriptsize 0},\ldots,\text{\scriptsize 0}}_{\text{\scriptsize n times}}}(x)
=  G\Biggl(\Biggl\{\underbrace{\frac{1}{\tau},\frac{1}{\tau},
  \cdots,\frac{1}{\tau}}_{\text{n times}}\Biggr\},x\Biggr)
=
\frac{1}{n!} \ln^n(x)~,
\end{equation}
see the appendix of Ref.~\cite{Ablinger:2017xml} for details. Square-root valued letters usually play a role in two mass 
OMEs but also for some single mass OMEs starting from three--loop order and related quantities
\cite{Ablinger:2014bra,Ablinger:2014yaa,Ablinger:2015tua,Ablinger:2017xml,Blumlein:2018jfm,Ablinger:2018brx,Blumlein:2019srk}.
\section{The massive operator matrix element}
\label{sec:4}

\vspace*{1mm}
\noindent
We obtain the following expression for the $O(\ep^0)$ term of the unrenormalized 3-loop two-mass pure singlet operator 
matrix element 
\begin{eqnarray}
\tilde{a}_{Qq}^{(3), \rm PS}(x) &=& C_F T_F^2 \Biggl\{
R_0(m_1,m_2,x) +\big(\theta(\eta_--x)+\theta(x-\eta_+)\big) x \, g_0(\eta,x)
\nonumber \\ &&
+\theta(\eta_+-x) \theta(x-\eta_-) \biggl[x \, f_0(\eta,x)
-\int_{\eta_-}^x dy \left(f_1(\eta,y)+\frac{x}{y} f_3(\eta,y)\right)\biggr]
\nonumber \\ &&
+\theta(\eta_--x) \int_x^{\eta_-} dy \left(g_1(\eta,y)+\frac{x}{y} g_3(\eta,y)\right)
\nonumber \\ &&
-\theta(x-\eta_+) \int_{\eta_+}^x dy \left(g_1(\eta,y)+\frac{x}{y} g_3(\eta,y)\right)
\nonumber \\ &&
+x \, h_0(\eta,x) +\int_x^1 dy \left(h_1(\eta,y)+\frac{x}{y} h_3(\eta,y)\right)
\nonumber \\ &&
+\theta(\eta_+-x) \int_{\eta_-}^{\eta_+} dy \left(f_1(\eta,y)+ \frac{x}{y} f_3(\eta,y)\right)
\nonumber \\ &&
+\int_{\eta_+}^1 dy \left(g_1(\eta,y)+\frac{x}{y} g_3(\eta,y)\right)
\Biggr\}.
\label{aQq}
\end{eqnarray}
Here we follow the notation used in Ref.~\cite{Ablinger:2017xml}. In the present case no functions carrying the index 2 occur.
The functions $g_i(\eta,x)$ in Eq.~(\ref{aQq}) shall not be confounded with polarized structure functions, also often 
denoted
by $g_i$. Here $\theta(z)$ denotes the Heaviside function
\begin{eqnarray}
\theta(z) = \left\{\begin{array}{ll} 1 &~~z \geq 0\\
                                     0 &~~z < 0.
\end{array} \right.
\end{eqnarray}
The pole terms are obtained in analytic form in terms of harmonic polylogarithms, cf. (\ref{Ahhhqq3PSQ}).
For convenience we define the auxiliary functions $u$ and $v$ as
\begin{equation}
u=\frac{x(1-x)}{\eta},~~~~
v=\frac{\eta}{x(1-x)}.
\end{equation}
If in the following expressions the harmonic polylogarithms $H_{\vec{a}}$ are given without argument it is understood that 
their 
argument is $x$.
The functions appearing in Eq.~(\ref{aQq}) are given by
\begin{eqnarray}
R_0(m_1,m_2,x)&=&32
        \bigg( L_1^3 +L_1  L_2  (L_1  +L_2  ) + L_2^3 \bigg)
\bigg[5 (-1+x) -2 (1+x) H_0 \bigg]\nonumber\\
& & +128 L_1  L_2  \bigg[
        (x+1) \bigg(\frac{2}{3}H_{0,1}-\frac{10}{9}H_0-\frac{2}{3}\zeta_2\bigg)
        +(x-1)\bigg(\frac{10}{9}-\frac{5}{3}H_1\bigg)
\bigg]\nonumber\\
&&+32 \big(
         L_1^2
        + L_2^2
\big)
\bigg[(x+1)\bigg(
		\frac{2}{3}  H_{0,1}
		+H_0^2
		-\frac{2}{3} \zeta_2
		\bigg)
		+(x-1)\bigg(
		\frac{1}{9}
        -\frac{5}{3} H_1
        \bigg)\nonumber\\
&&        +\frac{1}{9} (17-37 x) H_0
\bigg]\nonumber\\
&&+64 (L_1 +L_2 ) \bigg[
        (1+x) \bigg(
        \Big(
                2 H_{0,1}
                -\frac{8 \zeta_2}{3}
        \Big) H_0
        -\frac{2}{9} H_0^3
        -\frac{10}{3} H_{0,0,1}\nonumber\\
&&        -\frac{4}{3} H_{0,1,1}
        +\frac{14}{3} \zeta_3
\bigg)
+(x-1) \bigg(
        \frac{442}{27}
        +\frac{5}{3} H_1^2
        -\frac{5}{9} H_1 \Big(
                1+9 H_0\Big)
\bigg)\nonumber\\
&&-\frac{2}{27} (56+137 x) H_0
+\frac{1}{9} (-5+4 x) H_0^2
+\frac{2}{9} (-17+28 x) H_{0,1}\nonumber\\
&&+\frac{2}{9} (-28+17 x) \zeta_2
\bigg]
+\frac{64}{1215} \Bigg[(1+x) \bigg(
        \Big(
                3240 H_{0,0,1}
                +1620 H_{0,1,1}
        \Big) H_ 0\nonumber\\
&&        +\Big(
                -1620 H_{0,1}
                +945 \zeta_2
        \Big) H_ 0^2
        +90 H_ 0^4
        -1080 H_{0,0,0,1}\nonumber\\
&&        -2700 H_{0,0,1,1}
        +540 H_{0,1,1,1}
        +1296 \zeta_2^2
\bigg)+
(-1+x) \bigg(
        20 (437+54 x)
        +\nonumber\\
&&        \Big(
                1080 H_ 0
                +4050 H_ 0^2
                +2025 \zeta_2
        \Big) H_ 1
        -225 H_ 1^3\nonumber\\
&&        -45 H_ 1^2 \Big(
                11+45 H_ 0\Big)
\bigg)
+\Big(
        -10 \big(
                -842+1111 x+81 x^2\big)\nonumber\\
&&        -540 (-7+11 x) H_{0,1}
        -45 (-53+73 x) \zeta_2
        -4860 (1+x) \zeta_3
\Big) H_ 0\nonumber\\
&&+165 (19+37 x) H_ 0^2
-30 (-19+8 x) H_ 0^3
+30 (-1+x) (157+27 x) H_ 1\nonumber\\
&&+\Big(
        -30 (61+169 x)
        -810 (1+x) \zeta_2
\Big) H_{0,1}
+180 (-11+25 x) H_{0,0,1}\nonumber\\
&&+180 (-14+13 x) H_{0,1,1}
+15 (131+329 x) \zeta_2
+90 (-55+29 x) \zeta_3
\Bigg],
\\
g_0(\eta,x)&=&
-\frac{32 (1-x)}{9} \Bigg[
        -\frac{16 (-1+x) x}{\eta }
        +18 \bigg(
                -\frac{2 (\eta 
                -4 (-1+x) x
                )^2}{9 \eta ^2}
                +\frac{1}{3} \zeta_2
        \bigg) H_0\big(
                u\big)
\nonumber\\
&&        +5 H_0^2\big(
                u\big)
        +2 \bigg(
                -1
                +\frac{(\eta 
                -4 (1-x) x
                )^{3/2}}{\eta ^{3/2}}
        \bigg) \zeta_2
        +H_0^3\big(
                u\big)
\Bigg]\nonumber\\
&&-\frac{64 (1-x)}{9} \Bigg[
        \bigg(
                2 \frac{(\eta 
                -4 (1-x) x
                )^{3/2}}{\eta ^{3/2}}
                -3 \zeta_2
        \bigg) G\left(
                \left\{\frac{\sqrt{1-4 \tau }}{\tau }\right\},u\right)\Bigg]\nonumber\\
&&+\frac{64 (1-x)}{3} \left[ 
        G^2\left(
                \left\{\frac{\sqrt{1-4 \tau }}{\tau }\right\},u\right)
+
        G\left(
                \left\{\frac{\sqrt{1-4 \tau }}{\tau },\frac{\sqrt{1-4 \tau \
}}{\tau },\frac{1}{\tau }\right\},u\right)\right]\nonumber\\
&&
-\frac{64 (1-x)}{9} 
        G\left(\left\{
                \frac{\sqrt{1-4 \tau }}{\tau },\frac{1}{\tau \
}\right\},u\right)
\frac{(\eta 
-4 (1-x) x
)^{3/2}}{\eta ^{3/2}},\\
g_1(\eta,x)&=&\frac{64}{27 \eta ^2 x} \bigg[
        -6 (1-x) H_ 0\big(
                u\big) P_ 2
        -8 \eta  (-1+x) (1+x) (7 \eta 
        +24 (1-x) x
        )\nonumber\\
&&        +3 \eta ^2 (-1+x) (-5+13 x) H_ 0^2\big(
                u\big)
        -3 \eta ^2 (1-x) (-1+2 x) H_ 0^3\big(
                u\big)\nonumber\\
&&        -(6 (1-x)) \bigg(
                (1+x) \eta ^{3/2}
                -4 \eta  (1+x) \sqrt{\eta 
                -4 (1-x) x
                }
                +2 (-1+x) x (1+10 x) \nonumber\\
&&           \times     \sqrt{\eta 
                -4 (1-x) x
                }
        \bigg) \sqrt{\eta } \zeta_2
\bigg]
+\frac{128 (-1+x)}{9 x} \bigg[
        \Big(
                -4 \frac{\sqrt{\eta 
                -4 (1-x) x
                }}{\eta ^{3/2}} P_ 1\nonumber\\
&&                -3 (-1+2 x) \zeta_2
        \Big) G\left(
                \left\{\frac{\sqrt{1-4 \tau }}{\tau }\right\},u\right)\bigg]\nonumber\\
&&-\frac{128 (-1+x) (-1+2 x)}{3 x} 
        G^2\left(
                \left\{\frac{\sqrt{1-4 \tau }}{\tau }\right\},u\right)\nonumber\\
&&-\frac{128 (-1+x) (-1+2 x)}{3 x} 
       G\left(
                \left\{\frac{\sqrt{1-4 \tau }}{\tau },\frac{\sqrt{1-4 \tau \
}}{\tau },\frac{1}{\tau }\right\},u\right)\nonumber\\
&&-\frac{256 (-1+x) P_ 1}{9 x} 
        G\left(
                \left\{\frac{\sqrt{1-4 \tau }}{\tau },\frac{1}{\tau \
}\right\},u\right) 
\frac{\sqrt{\eta 
-4 (1-x) x
}
}{\eta ^{3/2}},\\
g_3(\eta,x)&=&-\frac{32}{27 \eta ^2 x} \bigg[
        8 \eta  (1-x) P_4
        -6 (1-x) H_0\big(
                u\big) P_5
        +3 \eta ^2 (-1+x) (-5+8 x)\nonumber\\
&&         \times H_0^2\big(
                u\big)
        +3 \eta ^2 (-1+x)^2 H_0^3\big(
                u\big)
        -(6 (1-x)) \bigg(
                (1+2 x) \eta ^{3/2}\nonumber\\
&&                -\eta  (4+5 x) \sqrt{\eta 
                -4 (1-x) x
                }
                +2 (-1+x) x (1+8 x) \sqrt{\eta 
                -4 (1-x) x
                }
        \bigg) \sqrt{\eta } \zeta_2
\bigg]\nonumber\\
&&-\frac{64 (1-x)}{9 x} \bigg[
        \bigg(
                2 \frac{\sqrt{\eta 
                -4 (1-x) x
                }}{\eta ^{3/2}} P_3
                +3 (-1+x) \zeta_2
        \bigg) G\left(
                \left\{\frac{\sqrt{1-4 \tau }}{\tau }\right\},u\right)\bigg]\nonumber\\
&&+\frac{64 (-1+x)^2}{3 x} 
        G^2\left(
                \left\{\frac{\sqrt{1-4 \tau }}{\tau }\right\},u\right)\nonumber\\
&&+\frac{64 (-1+x)^2}{3 x} 
        G\left(
                \left\{\frac{\sqrt{1-4 \tau }}{\tau },\frac{\sqrt{1-4 \tau \
}}{\tau },\frac{1}{\tau }\right\},u\right)\nonumber\\
&&-\frac{64 (1-x) P_3}{9 x} 
        G\left(
                \left\{\frac{\sqrt{1-4 \tau }}{\tau },\frac{1}{\tau \
}\right\},u\right) \frac{\sqrt{\eta 
-4 (1-x) x
}}{\eta ^{3/2}},
\end{eqnarray}
with the polynomials
\begin{eqnarray}
P_1&=&2 \eta  (x+1)-10 x^3+9 x^2+x,\\
P_2&=&3 \eta ^2 (2 x \zeta_2+x-\zeta_2+1)+8 \eta  x \left(10 x^2-9 \
x-1\right)-16 (1-x)^2 x^2 (10 x+1),\\
P_3&=&\eta  (5 x+4)+2 x \left(-8 x^2+7 x+1\right),\\
P_4&=&7 \eta  (x+1)+6 x \left(-5 x^2+x+4\right),\\
P_5&=&\eta ^2 (x (3 \zeta_2+5)-3 \zeta_2+3)+8 \eta  x \left(8 x^2-7 \
x-1\right)-16 (x-1)^2 x^2 (8 x+1),
\end{eqnarray}
and
\begin{eqnarray}
f_0(\eta,x)&=&\bigg[
        -
        \frac{16 (1-x)}{3} 
                G\left(
                        \left\{\frac{1}{\tau },\sqrt{4-\tau } \sqrt{\tau \
}\right\},v\right)
        -\frac{4 P_6}{9 (-1+x) x^2} \Big[
                -1
                +2 H_0\big(
                        v\big)\Big] \nonumber\\
&&                       \times \frac{(-\eta 
        +4 (1-x) x
        )^{3/2}}{\eta ^{3/2}}
\bigg] G\left(
        \left\{\sqrt{4-\tau } \sqrt{\tau }\right\},v\right)\nonumber\\
&&+\frac{4 (1-x)}{3} \bigg[
        \Big[
                -1+2 H_0\big(
                        v\big)\Big] G^2\left(
                \left\{\sqrt{4-\tau } \sqrt{\tau }\right\},v\right)\bigg]\nonumber\\
&&+\frac{16 (1-x)}{3} 
        G\left(
                \left\{\frac{1}{\tau },\sqrt{4-\tau } \sqrt{\tau },\sqrt{4-\tau } \sqrt{\tau }\right\},v\right)
+\frac{1}{18} \bigg[
        -1536 (1-x)\nonumber\\
&&        -\frac{9 \eta ^4}{(-1+x)^3 x^4}
        -\frac{80 \eta ^3}{(-1+x)^2 x^3}
        -\frac{104 \eta ^2}{(-1+x) x^2}
        +\frac{576 \eta }{x}
        +\frac{4  P_7}{(-1+x)^3 x^4}H_0\big(
                v\big)\nonumber\\
&&        -320 (1-x) H_0^2\big(
                v\big)
        +64 (1-x) H_0^3\big(
                v\big)
        -\Big(128 (1-x)\Big) \Big(
                5
               -3 H_0\big(
                        v\big)\Big) \zeta_2\nonumber\\
&&        +768 (1-x) \zeta_3
\bigg]
-\frac{8 P_6}
{9 (1-x) x^2} 
        G\left(
                \left\{\frac{1}{\tau },\sqrt{4-\tau } \sqrt{\tau \
}\right\},v\right) \nonumber\\
&&\times\frac{(-\eta 
+4 (1-x) x
)^{3/2}}{\eta ^{3/2}},\\
f_1(\eta,x)&=&\frac{1}{27 x^5} \Bigg[
        -\frac{1}{(1-x)^3} \bigg[
                6912 \eta  (-1+x)^3 x^4
                -27 \eta ^4 (-1+2 x)
                -240 \eta ^3 (-1+x) x \nonumber\\
&&                (-1+2 x)
                +512 (-1+x)^4 x^4 (-16+11 x)
                +12 H_ 0\big(
                        v\big) P_ 9\nonumber\\
&&                -24 \eta ^2 (-1+x)^2 x^2 (-25+2 x)
                +192 (-1+x)^4 x^4 (-5+13 x) H_ 0^2\big(
                        v\big)\nonumber\\
&&                -192 (-1+x)^4 x^4 (-1+2 x) H_ 0^3\big(
                        v\big)
        \bigg]
        +384 (1-x) x^4 \bigg(
                5
                -13 x\nonumber\\
&&                +3 (-1+2 x) H_ 0\big(
                        v\big)
        \bigg) \zeta_2
        -2304 (-1+x) x^4 (-1+2 x) \zeta_3
\Bigg]
+\nonumber\\
&&\bigg[
        \frac{32 (-1+x) (-1+2 x)}{3 x} 
                G\left(
                        \left\{\frac{1}{\tau },\sqrt{4-\tau } \sqrt{\tau \
}\right\},v\right)\nonumber\\
&&        -\frac{8 P_ 8}{9 (1-x) x^3} \bigg(
                -1+2 H_ 0\big(
                        v\big)\bigg) \frac{\sqrt{-\
\eta 
        +4 (1-x) x
        }}{\eta ^{3/2}}
\bigg] \nonumber\\
&& \times G\left(
        \left\{\sqrt{4-\tau } \sqrt{\tau }\right\},v\right)
-\frac{8 (-1+x) (-1+2 x)}{3 x} \bigg[
        \bigg(
                -1+2 H_ 0\big(
                        v\big)\bigg) \nonumber\\
&& \times G^2\left(
                \left\{\sqrt{4-\tau } \sqrt{\tau }\right\},v\right)\bigg]
+\frac{32 (1-x) (-1+2 x)}
{3 x} \nonumber\\
&& 
        \times G\left(
                \left\{\frac{1}{\tau },\sqrt{4-\tau } \sqrt{\tau },\sqrt{4-\tau } \sqrt{\tau }\right\},v\right)
-\frac{16 P_ 8}{9 (-1+x) x^3} \nonumber\\
&&        \times G\left(
                \left\{\frac{1}{\tau },\sqrt{4-\tau } \sqrt{\tau \
}\right\},v\right) \frac{\sqrt{-\eta 
+4 (1-x) x
}}{\eta ^{3/2}},\\
f_3(\eta,x)&=&\frac{1}{54 (-1+x)^2 x^5} \bigg[
        27 \eta ^4
        -240 \eta ^3 (1-x) x
        -5184 \eta  (-1+x)^2 x^4
        -1024 (-8+x) \nonumber\\
&&        \times (-1+x)^3 x^4
        -12 H_ 0\big(
                v\big) P_{10}
        -24 \eta ^2 (-1+x) x^2 (25+11 x)\nonumber\\
&&        -192 (-1+x)^3 x^4 (-5+8 x) H_ 0^2\big(
                v\big)
        +192 (-1+x)^4 x^4 H_ 0^3\big(
                v\big)\nonumber\\
&&        +384 (-1+x)^3 x^4 \bigg(
                5
                -8 x
                -3 (1-x) H_ 0\big(
                        v\big)
        \bigg) \zeta_2
        +2304 (-1+x)^4 x^4 \zeta_3
\bigg]\nonumber\\
&&+\bigg[
        -\frac{16 (-1+x)^2}{3 x}
                G\left(
                        \left\{\frac{1}{\tau },\sqrt{4-\tau } \sqrt{\tau \
}\right\},v\right)
        -\frac{4 P_{11}}{9 x^3} \bigg(
                -1\nonumber\\
&&                +2 H_ 0\big(
                        v\big)\bigg) \frac{\sqrt{-\
\eta 
        +4 (1-x) x
        }}{\eta ^{3/2}}
\bigg] G\left(
        \left\{\sqrt{4-\tau } \sqrt{\tau }\right\},v\right)\nonumber\\
&&+\frac{4 (-1+x)^2}{3 x} 
\bigg[
        \bigg(
                -1+2 H_ 0\big(
                        v\big)\bigg) G\left(
                \left\{\sqrt{4-\tau } \sqrt{\tau }\right\},v\right)^2\bigg] \nonumber\\
&&+\frac{16 (-1+x)^2}{3 x} 
        G\left(
                \left\{\frac{1}{\tau },\sqrt{4-\tau } \sqrt{\tau },\sqrt{4-\tau } \sqrt{\tau }
                \right\},v\right)
+\frac{8 P_{11}}{9 x^3} \nonumber\\
&& \times       G\left(
                \left\{\frac{1}{\tau },\sqrt{4-\tau } \sqrt{\tau \
}\right\},v\right) \frac{\sqrt{-\eta 
+4 (1-x) x
}}{\eta ^{3/2}},
\end{eqnarray}
with
\begin{eqnarray}
P_6    &=&  3 \eta ^2+6 \eta  (1-x) x+4 (x-1)^2 x^2,\\
P_7    &=&  3 \eta ^4-24 \eta ^3 (1-x) x+20 \eta ^2 (x-1)^2 x^2-160 \eta  \
            (x-1)^3   x^3-128 (x-1)^4 x^4,\\
P_8    &=&  \eta ^3 (6 x-3)+6 \eta ^2 x \left(2 x^2-3 x+1\right)-8 \eta  \
            (x-1)^2 x^2 (8 x-1)\nonumber\\
       & &  +8 (x-1)^3 x^3 (10 x+1),\\
P_9    &=&  \eta ^4 (6 x-3)+24 \eta ^3 x \left(2 x^2-3 x+1\right)-4 \eta ^2 (x-1)^2 x^2 (2 x+11)\nonumber\\
       & &  -64 \eta  (x-1)^3 x^3 (8 x-1)+96 (x-1)^4 x^4 (x+4),\\
P_{10} &=&  3 \eta ^4-24 \eta ^3 (1-x) x-4 \eta ^2 x^2 \left(7 x^2+4 x-11\right)-32 \eta  (x-1)^2 x^3 (11 x-2)\nonumber\\
       & &  +32 (x-1)^3 x^4 (7 x+12),\\
P_{11} &=&  3 \eta ^3-6 \eta ^2 (1-x) x-4 \eta  x^2 \left(11 x^2-13 x+2\right)+8 (1-x)^2 x^3 (8 x+1).
\end{eqnarray}
The functions $h_i$ are defined as follows 
\begin{eqnarray}
h_i(\eta,x) = g_i\left(\frac{1}{\eta},x\right), \quad i=0,1,3.
\end{eqnarray}
In deriving the expressions given above, we have used shuffle algebra relations wherever possible, cf.~\cite{Blumlein:2003gb}.
The function $R_0(m_1,m_2,x)$ arises from the residues taken in order to resolve the singularities in $\ep$ of the 
contour integrals in the functions $B_i$.
The functions $f_i(\eta,x)$, $g_i(\eta,x)$, with $i=0,1,3$, arise from the sum of residues
of the contour integrals that remain after the $\ep$ expansion, as described in the previous section. The functions with 
$i=0$ are those where no additional factor depending on $N$ occurs. The functions with $i=1$, and $i=3$ 
are those where a factor of $1/N$ and $1/(N+1)$ was absorbed, respectively, see Eqs.~(\ref{absorbN1}, \ref{absorbN2}). 
The different Heaviside functions restrict the corresponding values of $x$ to the appropriate regions.

Since no contour integral needs to be performed in the case $R_0(m_1,m_2,x)$, the easiest way to compute this function 
is to integrate in $x$ and then perform the Mellin inversion using {\tt HarmonicSums}. The expressions of the 
$G$-functions in the above equations can all be given  in terms of harmonic polylogarithms, cf. Appendix of 
Ref.~\cite{Ablinger:2017xml}, containing  square-root valued arguments. 

We see that iterated integrals of up to weight three appear in our result. The alphabet of these integrals is given 
in terms of just three letters:
\begin{equation}
\frac{1}{\tau}, \quad \sqrt{4-\tau} \sqrt{\tau}, \quad \frac{\sqrt{1-4 \tau}}{\tau}.
\end{equation}
One may try to integrate the remaining integrals in Eq.~(\ref{aQq}) over $y$ into iterated integrals of 
higher weight. However, the numerical representation needs to be done in addition, unlike the case of 
harmonic polylogarithms \cite{Gehrmann:2001pz,Ablinger:2018sat}.

In order to compute the corresponding contribution to the structure function
$g_1(x,Q^2)$ or for the transition rate in the VFNS, we have to perform the   
convolution with parton distribution functions, which can be obtained straightforwardly.
\section{Numerical Results}
\label{sec:5}

\begin{figure}[H]\centering
\includegraphics[width=0.7\textwidth]{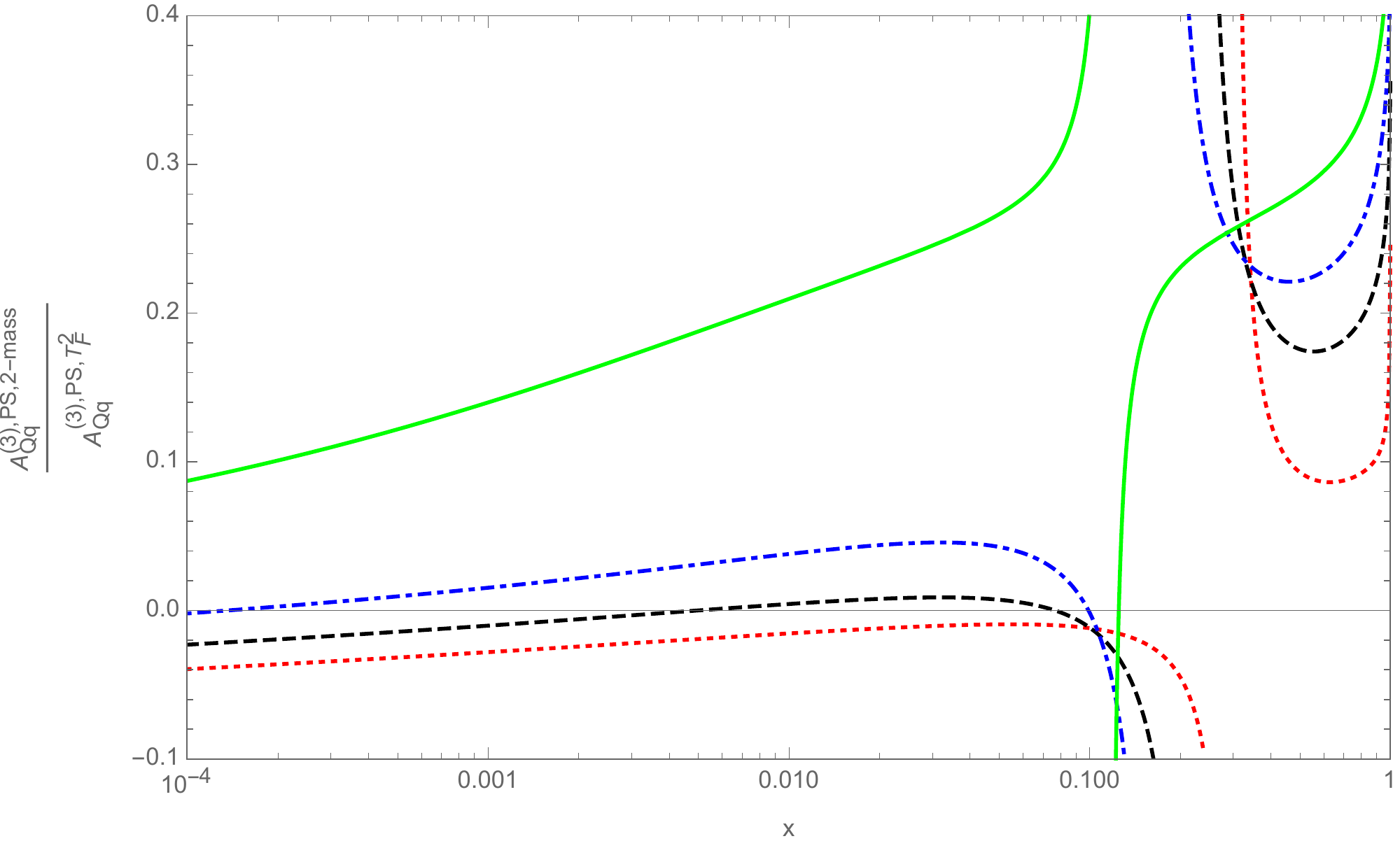}
\caption[]{\label{FIGnum} \sf
The ratio of the 2-mass (tm) contributions to the massive OME $A_{Qq}^{{\sf PS},(3)}$ to all contributions to
$A_{Qq}^{{\sf PS},(3)}$
of $O(T_F^2)$ as a function of $x$ and $\mu^2$.
Dotted line (red): $\mu^2 = 30~\GeV^2$.
Dashed line (black): $\mu^2 = 50~\GeV^2$.
Dash-dotted line (blue): $\mu^2 = 100~\GeV^2$.
Full line (green): $\mu^2 = 1000~\GeV^2$. Here the on-shell heavy quark masses $m_c = 1.59~\GeV$ and $m_b =
4.78~\GeV$
\cite{Alekhin:2012vu,Agashe:2014kda} have been used.
}
\end{figure}

\vspace*{1mm}
\noindent
We compare the polarized pure singlet 2-mass contributions to the  complete $O(T_F^2 C_{F})$ term  
as a function of $x$ and $\mu^2$ in Figure~\ref{FIGnum}.
Typical virtualities are $\mu^2 \in [30,1000]~\GeV^2$. The ratio of the 2-mass contributions to the complete term 
of $O(T_F^2 C_{F})$ has a singularity around $x \sim 0.1$. At lower virtualities the corrections are nearly 
constant in the small $x$ region and grow with $\mu^2$ rising from negative to positive values. In the large $x$ region 
the 
ratio falls and rises once again towards $x \rightarrow 1$. At $\mu^2 = 1000~\GeV^2$ the corrections are comparatively 
large 
and positive due to the large logarithms, except in the pole region. In size the corrections are comparable to those found 
in the unpolarized case and do majorly range between $-0.1$ to $0.4$.
\section{Conclusions}
\label{sec:6}

\vspace*{1mm}
\noindent
We have calculated the two-mass 3-loop contributions to the polarized massive OME $A_{Qq}^{{\sf PS},(3)}$ in analytic 
form in $x$-space for a general mass-ratio $\eta$ in the Larin scheme. It contributes to the massive 3-loop Wilson
coefficient of the deep-inelastic structure function $g_1(x,Q^2)$ in the region $m^2 \ll Q^2$ and is, as well, one
of the polarized OMEs in the two--mass 3-loop VFNS, needed to describe the process of heavy quarks becoming massless 
at large virtualities. As a function of $x$, its relative contribution to the $O(T_F^2 C_{A,F})$ terms of the whole 
matrix element $A_{Qq}^{{\sf PS, TF^2},(3)}$ lay in the region of about $[-0.1, 0.4]$ and exhibit a pole at $x \sim 0.1$.
the two-mass contribution is not negligible against the single mass contributions.

We applied Mellin-Barnes techniques to obtain the $x$-space result by factoring out the $N$-dependence in terms of the 
kernel $x^N$, and used integration by parts to absorb the $N$-dependent polynomial pre-factors. The result can be 
written as single limited integrals within the range $x \in [0,1]$ over iterated integrals containing also square-root 
valued letters. These integrals can be turned into polylogarithms of involved root-valued arguments, 
depending on the real parameter $\eta$. The odd Mellin moments of the OME exhibit a growing number of polynomial terms 
in $\eta$ with growing values of $N$. Due to this structural property and the arbitrariness of $\eta$, which enters 
the ground field, the method of arbitrarily large moments \cite{Blumlein:2017dxp} cannot be used to find the result 
in the present case. The set of necessary integrals for these representations has already been derived in the 
unpolarized case in Ref.~\cite{Ablinger:2017xml}. The concept of square root-valued iterated integrals turned out
to be of central importance in deriving the present results. Moreover, their weights are such that one can still relate 
them to harmonic polylogarithms of more complicated arguments.\footnote{This has been a principle also in early 
Mellin-representations of harmonic sums \cite{Blumlein:1998if} and \cite{Blumlein:2009ta}, limiting the alphabet 
to that of Nielsen integrals \cite{NIELS}.}   

The Larin scheme is one of the valid schemes to perform calculations in the polarized case. At present the massless 
polarized three-loop Wilson coefficients are not yet available \cite{VERM2}. They will also be calculated in the Larin 
scheme first. Together with parton distribution functions, evolved in the Larin scheme, one can then form observables like
$g_1(x,Q^2)$ and related quantities \cite{Blumlein:2004xs}. The anomalous dimensions for the Larin scheme are available to 
three--loop order \cite{Moch:2014sna,Behring:2019tus}.


\vspace*{5mm}
\noindent
{\bf Acknowledgment.}\\
This work was supported in part by the Austrian Science Fund (FWF) grant SFB F50 (F5009-N15) and has received funding 
from the European Union's Horizon 2020 research and innovation programme under the Marie Sk\/{l}odowska-Curie grant  
agreement No. 764850, SAGEX, and COST action CA16201: Unraveling new physics at the LHC through the precision 
frontier.
\newpage
\appendix

\section{Fixed Moments of \boldmath $\tilde{A}_{Qq}^{\text{PS, tm},(3)}$}
\label{app1}

\noindent
For fixed values of $N = 2k+1, k \in \mathbb{N}$, the difference equations
factorize to first order and we find the following moments, unexpanded in
the mass ratio $\eta$. In the following we will use the notation
\begin{eqnarray}
      H_{\vec{w}} (\sqrt{\eta}) &\equiv& H_{\vec{w}}.
\end{eqnarray}
The fixed moments are given by      
\begin{eqnarray}
\lefteqn{\hat{\tilde{A}}_{Qq}^{\text{PS, tm},(3)}(N=1) =} \nonumber \\ &&
C_F T_F^2 \Biggl\{
        \frac{256}{3 \ep^2}
        +\frac{1}{\ep}
        \biggl[
            \frac{320}{9}
            +64 \bigl( L_1 + L_2 \bigr)
        \biggr]
        -\frac{3}{2 \eta ^{3/2}} 
        \biggl[
                -2 Q_{1} 
                + Q_{2} \bigl( H_1 + H_{-1} \bigr)
        \biggr] 
\nonumber \\ &&
      \times
            \biggl( L_1^2 + L_2^2 \biggr)
        + \frac{3}{\eta^{3/2}} L_1 L_2 
        \biggl[
                        -2 Q_{3}
                        + Q_{2} \bigl( H_1 + H_{-1} \bigr)
        \biggr] 
\nonumber \\ &&
        -\frac{2}{3 \eta ^{3/2}} 
        \biggl[                
                        2 Q_{4} 
                        +9 Q_{2} \bigl( H_{0,1} + H_{0,-1} \bigr)
        \biggr] L_1
\nonumber \\ &&
        +\frac{2}{3 \eta ^{3/2}} 
        \biggl[
                2 Q_{5} 
                +9 Q_{2} \bigl( H_{0,1} + H_{0,-1} \bigr)
        \biggr] L_2
\nonumber \\ &&
        -\frac{4}{27 \eta ^{3/2}} 
        \biggl[
                -2 Q_{6} 
                +81 Q_{2} \bigl( H_{0,0,1} + H_{0,0,-1} \bigr)
        \biggr] 
        +32 \zeta_2,
\end{eqnarray}
\begin{eqnarray}
Q_{1} &=& \sqrt{\eta } \bigl( 1+10 \eta +\eta ^2 \bigr),
\\
Q_{2} &=& (1-\eta )^2 (1+\eta ),
\\
Q_{3} &=& \sqrt{\eta } \bigl( 1-6 \eta +\eta ^2 \bigr),
\\
Q_{4} &=& \sqrt{\eta } \bigl( -9-20 \eta +9 \eta ^2 \bigr),
\\
Q_{5} &=& \sqrt{\eta } \bigl( -9+20 \eta +9 \eta ^2 \bigr),
\\
Q_{6} &=& \sqrt{\eta } \bigl( 81+284 \eta +81 \eta ^2 \bigr),
\end{eqnarray}
\begin{eqnarray}
\lefteqn{\hat{\tilde{A}}_{Qq}^{\text{PS, tm},(3)}(N=3) =} \nonumber \\ &&
C_F T_F^2 \Biggl\{
        -\frac{1280}{81 \ep^3}
        +\frac{1}{\ep^2}
        \biggl[
            \frac{1760}{243}
            -\frac{320}{27} \bigl( L_1 + L_2 \bigr)
        \biggr] 
        +\frac{1}{\ep}
        \biggl[
           -\frac{12820}{729}
\nonumber \\ &&
           -\frac{160}{27} \bigl( L_1^2 + L_1 L_2 + L_2^2 \bigr)
           +\frac{440}{81} \bigl( L_1 + L_2 \bigr)
          -\frac{160}{27} \zeta_2
        \biggr]
\nonumber \\ &&
        -\frac{200}{81} L_1^3
        -\frac{160}{81} L_2^3
        -\frac{80}{27} L_1 L_2^2
        -\frac{40}{27} L_2 L_1^2
        + (L_1^2 + L_2^2) \biggl[
                \frac{5 Q_{7}}{324 \eta }
\nonumber \\ &&
                -\frac{5 Q_{12}}{216} \frac{H_1}{\eta ^{3/2}}
                -\frac{5 Q_{13}}{216} \frac{H_{-1}}{\eta ^{3/2}}
        \biggr]
                +L_1 L_2 \biggl[
                        -\frac{5 Q_{8}}{162 \eta }
\nonumber \\ &&
                        +\frac{5 Q_{12}}{108} \frac{H_1}{\eta ^{3/2}}
                        +\frac{5 Q_{13}}{108} \frac{H_{-1}}{\eta ^{3/2}}
                \biggr]
        + L_1 \biggl[
                -\frac{5 Q_{9}}{243 \eta }
\nonumber \\ &&
                -\frac{5 Q_{12}}{54}
                         \frac{H_{0,1}}{\eta ^{3/2}}
                -\frac{5 Q_{13}}{54} 
                         \frac{H_{0,-1}}{\eta ^{3/2}}
                -\frac{40}{9} \zeta_2
        \biggr]
        +\frac{5 Q_{10}}{4374 \eta }
\nonumber \\ &&
        +L_2 \biggl[
                \frac{5 Q_{11}}{243 \eta }
                +\frac{5 Q_{12}}{54} 
                         \frac{H_{0,1}}{\eta ^{3/2}}
                +\frac{5 Q_{13}}{54} 
                         \frac{H_{0,-1}}{\eta ^{3/2}}
                -\frac{40}{9} \zeta_2
        \biggr]
\nonumber \\ &&
        -\frac{5 Q_{12}}{27} 
                 \frac{H_{0,0,1}}{\eta ^{3/2}}
        -\frac{5 Q_{13}}{27}
                 \frac{H_{0,0,-1}}{\eta ^{3/2}}
        +\frac{220}{81} \zeta_2
        +\frac{160}{81} \zeta_3
\Biggr\}
\end{eqnarray}
\begin{eqnarray}
Q_{7} &=& 15+262 \eta +15 \eta ^2,
\\
Q_{8} &=& 15-2 \eta +15 \eta ^2,
\\
Q_{9} &=& -45+641 \eta +45 \eta ^2,
\\
Q_{10} &=& 1620+5659 \eta +1620 \eta ^2,
\\
Q_{11} &=& -45-641 \eta +45 \eta ^2,
\\
Q_{12} &=& 5+27 \eta +27 \eta ^2+5 \eta ^3-64 \eta ^{3/2},
\\
Q_{13} &=& 5+27 \eta +27 \eta ^2+5 \eta ^3+64 \eta ^{3/2},
\end{eqnarray}
\begin{eqnarray}
\lefteqn{\hat{\tilde{A}}_{Qq}^{\text{PS, tm},(3)}(N=5) =} \nonumber \\ &&
C_F T_F^2 \Biggl\{
        -\frac{14336}{2025 \ep^3}
        +\frac{1}{\ep^2}
      \biggl[
         \frac{20608}{6075}
        -\frac{3584}{675} \bigl( L_1 + L_2 \bigr)
      \biggr]
        +\frac{1}{\ep}
      \biggl[
        -\frac{3724784}{455625}
\nonumber \\ &&
        -\frac{1792}{675} \bigl( L_1^2 + L_1 L_2 + L_2^2 \bigr)
        +\frac{5152}{2025} \bigl( L_1 + L_2 \bigr)
        -\frac{1792}{675} \zeta_2
      \biggr]
        -\frac{448}{405} L_1^3
\nonumber \\ &&
        -\frac{1792}{2025} L_2^3
        -\frac{896}{675} L_1 L_2^2
        -\frac{448}{675} L_1^2 L_2
        +\frac{7 Q_{16}}{583200 \eta ^2} 
\nonumber \\ &&
        + (L_1^2 + L_2^2) \biggl[                
                 \frac{7 Q_{17} }{864000 \eta ^2} 
                -\frac{7 Q_{14} }{345600 \eta ^{5/2}} H_1
                -\frac{7 Q_{15} }{345600 \eta ^{5/2}} H_{-1}
        \biggr]
\nonumber \\ &&
                + L_1 L_2 \biggl[
                        -\frac{7 Q_{18} }{432000 \eta ^2} 
                        +\frac{7 Q_{14} }{172800 \eta ^{5/2}} H_1
                        +\frac{7 Q_{15} }{172800 \eta ^{5/2}} H_{-1}
                \biggr]
\nonumber \\ &&
        + L_1 \biggl[
                -\frac{7 Q_{19} }{9720000 \eta ^2} 
                -\frac{7 Q_{14} }{86400 \eta ^{5/2}} H_{0,1}
                -\frac{7 Q_{15} }{86400 \eta ^{5/2}} H_{0,-1}
                -\frac{448}{225} \zeta_2
        \biggr]
\nonumber \\ &&
        + L_2 \biggl[
                \frac{7 Q_{20} }{9720000 \eta ^2} 
                +\frac{7 Q_{14} }{86400 \eta ^{5/2}} H_{0,1}
                +\frac{7 Q_{15} }{86400 \eta ^{5/2}} H_{0,-1}
                -\frac{448}{225} \zeta_2
        \biggr]
\nonumber \\ &&
        -\frac{7 Q_{14} }{43200 \eta ^{5/2}} H_{0,0,1}
        -\frac{7 Q_{15} }{43200 \eta ^{5/2}} H_{0,0,-1}
        +\frac{2576 \zeta_2}{2025}
        +\frac{1792 \zeta_3}{2025}
\Biggr\},
\end{eqnarray}
\begin{eqnarray}
Q_{14} &=& 189+2425 \eta +13770 \eta ^2+13770 \eta ^3+2425 \eta ^4+189 \eta ^5-32768 \eta ^{5/2}, \\
Q_{15} &=& 189+2425 \eta +13770 \eta ^2+13770 \eta ^3+2425 \eta ^4+189 \eta ^5+32768 \eta ^{5/2}, \\
Q_{16} &=& 5103+65664 \eta +260834 \eta ^2+65664 \eta ^3+5103 \eta ^4,
\\
Q_{17} &=& 945+12440 \eta +230094 \eta ^2+12440 \eta ^3+945 \eta ^4,
\\
Q_{18} &=& 945+12440 \eta -5426 \eta ^2+12440 \eta ^3+945 \eta ^4, 
\\
Q_{19} &=& -42525-550350 \eta +8513792 \eta ^2+550350 \eta ^3+42525 \eta ^4,
\\
Q_{20} &=& -42525-550350 \eta -8513792 \eta ^2+550350 \eta ^3+42525 \eta ^4,
\end{eqnarray}
\begin{eqnarray}
\lefteqn{\hat{\tilde{A}}_{Qq}^{\text{PS, tm},(3)}(N=7) =} \nonumber \\ &&
C_F T_F^2 \Biggl\{
        -\frac{192}{49 \ep^3}
        +\frac{1}{\ep^2} \biggl[
         \frac{508}{245}
        -\frac{144 }{49} \bigl( L_1 + L_2 \bigr)
        \biggr]
        +\frac{1}{\ep} \biggl[
        -\frac{45155}{9604}
\nonumber \\ &&
        -\frac{72 }{49} \bigl( L_1^2 + L_1 L_2 + L_2^2 \bigr)
        +\frac{381 }{245} \bigl( L_1 + L_2 \bigr)
        -\frac{72}{49} \zeta_2
        \bigr]
        -\frac{30}{49} L_1^3
        -\frac{24}{49} L_2^3
\nonumber \\ &&
        -\frac{Q_{21}}{2765952000 \eta ^3} 
        -\frac{36}{49} L_1 L_2^2
        -\frac{18}{49} L_1^2 L_2
        + (L_1^2 + L_2^2) \biggl[                
                -\frac{3 Q_{22} }{14049280 \eta ^3} 
\nonumber \\ &&
                +\frac{9 Q_{23} }{802816 \eta ^{7/2}} H_1
                +\frac{9 Q_{24} }{802816 \eta ^{7/2}} H_{-1}
        \biggr]
        + L_1 L_2 \biggl[
             \frac{3 Q_{27}}{1404928 \eta ^3}
\nonumber \\ &&
            -\frac{9 Q_{23}}{401408 \eta ^{7/2}} H_1
            -\frac{9 Q_{24}}{401408 \eta ^{7/2}} H_{-1}
        \biggr]
        +L_2 \biggl[
                \frac{Q_{25}}{24586240 \eta ^3} 
\nonumber \\ &&
                -\frac{9 Q_{23} }{200704 \eta ^{7/2}} H_{0,1}
                -\frac{9 Q_{24} }{200704 \eta ^{7/2}} H_{0,-1}
                -\frac{54}{49} \zeta_2
        \biggr]
\nonumber \\ &&
        +L_1 \biggl[
                -\frac{Q_{26}}{24586240 \eta ^3} 
                +\frac{9 Q_{23} }{200704 \eta ^{7/2}} H_{0,1}
                +\frac{9 Q_{24} }{200704 \eta ^{7/2}} H_{0,-1}
                -\frac{54}{49} \zeta_2
        \biggr]
\nonumber \\ &&
        +\frac{9 Q_{23} }{100352 \eta ^{7/2}} H_{0,0,1}
        +\frac{9 Q_{24} }{100352 \eta ^{7/2}} H_{0,0,-1}
        +\frac{381}{490} \zeta_2
        +\frac{24}{49} \zeta_3
\Biggr\},
\end{eqnarray}
\begin{eqnarray}
Q_{21} &=& 22325625+170997750 \eta -1400033145 \eta ^2
-5593159388 \eta^3
\nonumber \\ &&
-1400033145 \eta ^4+170997750 \eta ^5+22325625 \eta ^6,
\\
Q_{22} &=& 4725+37590 \eta -284725 \eta ^2-5196076 \eta ^3-284725 \eta ^4
\nonumber \\ &&
+37590 \eta ^5+4725 \eta ^6,
\\
Q_{23} &=& 45+343 \eta -2835 \eta ^2-13937 \eta ^3-13937 \eta ^4-2835 \eta ^5+343 \eta ^6
\nonumber \\ &&
+45 \eta ^7+32768 \eta ^{7/2},
\\
Q_{24} &=& 45+343 \eta -2835 \eta ^2-13937 \eta ^3-13937 \eta ^4-2835 \eta ^5+343 \eta ^6
\nonumber \\ &&
+45 \eta ^7-32768 \eta ^{7/2},
\\
Q_{25} &=& 99225+767340 \eta -6163171 \eta ^2-86697600 \eta ^3+6163171 \eta ^4
\nonumber \\ &&
-767340 \eta ^5-99225 \eta ^6,
\\
Q_{26} &=& 99225+767340 \eta -6163171 \eta ^2+86697600 \eta ^3+6163171 \eta ^4
\nonumber \\ &&
-767340 \eta ^5-99225 \eta ^6,
\\
Q_{27} &=& 945+7518 \eta -56945 \eta ^2+53188 \eta ^3-56945 \eta ^4+7518 \eta ^5+945 \eta ^6,
\end{eqnarray}
\begin{eqnarray}
\lefteqn{\hat{\tilde{A}}_{Qq}^{\text{PS, tm},(3)}(N=9) =} \nonumber \\ &&
C_F T_F^2 \Biggl\{
        -\frac{45056}{18225 \ep^3}
        +\frac{1}{\ep^2} \biggl[
         \frac{2721664}{1913625}
        -\frac{11264}{6075} \bigl( L_1 + L_2 \bigr)
        \biggr]
\nonumber \\ &&
        +\frac{1}{\ep} \biggl[
        -\frac{5568605768}{1808375625}
        -\frac{5632}{6075} \bigl( L_1^2 + L_1 L_2 + L_2^2 \bigr)
        +\frac{680416}{637875} \bigl( L_1 + L_2 \bigr)
\nonumber \\ &&
        -\frac{5632 \zeta_2}{6075}
        \biggr]
        -\frac{1408}{3645} L_1^3
        -\frac{5632}{18225} L_2^3
        -\frac{2816}{6075} L_1 L_2 ^2
        -\frac{1408}{6075} L_1^2 L_2
\nonumber \\ &&
        +\frac{11 Q_{28}}{777746188800000 \eta ^4}
        + (L_1^2 + L_2^2) \biggl[                
                 \frac{11 Q_{29}}{62705664000 \eta ^4} 
                -\frac{11 Q_{30}}{398131200 \eta ^{9/2}} H_1
\nonumber \\ &&
                -\frac{11 Q_{31}}{398131200 \eta ^{9/2}} H_{-1}
        \biggr]
        + L_1 L_2 \biggl[
                        -\frac{11 Q_{32}}{4478976000 \eta ^4} 
                        +\frac{11 Q_{30}}{199065600 \eta ^{9/2}} H_1
\nonumber \\ &&
                        +\frac{11 Q_{31}}{199065600 \eta ^{9/2}} H_{-1}
                \biggr]
        +L_1
         \biggl[
                -\frac{11 Q_{33}}{4938071040000 \eta ^4} 
                -\frac{11 Q_{30}}{99532800 \eta ^{9/2}} H_{0,1}
\nonumber \\ &&
                -\frac{11 Q_{31}}{99532800 \eta ^{9/2}} H_{0,-1}
                -\frac{1408 \zeta_2}{2025}
        \biggr]
        +L_2 \biggl[
                \frac{11 Q_{34}}{4938071040000 \eta ^4} 
                +\frac{11 Q_{30}}{99532800 \eta ^{9/2}} H_{0,1}
\nonumber \\ &&
                +\frac{11 Q_{31}}{99532800 \eta ^{9/2}} H_{0,-1}
                -\frac{1408 \zeta_2}{2025}
        \biggr]
        -\frac{11 Q_{30}}{49766400 \eta ^{9/2}} H_{0,0,1}
\nonumber \\ &&
        -\frac{11 Q_{31}}{49766400 \eta ^{9/2}} H_{0,0,-1}
        +\frac{340208 \zeta_2}{637875}
        +\frac{5632 \zeta_3}{18225}
\Biggr\},
\end{eqnarray}
\begin{eqnarray}
Q_{28} &=& 84234583125+1142396443125 \eta -12238532911710 \eta ^2+27996557899275 \eta ^3
\nonumber \\ &&
+103736391394322 \eta ^4+27996557899275 \eta ^5-12238532911710 \eta ^6
\nonumber \\ &&
+1142396443125 \eta ^7+84234583125 \eta ^8, 
\\
Q_{29} &=& 848925+11764725 \eta -119776230 \eta ^2+247801995 \eta ^3+4258879634 \eta ^4
\nonumber \\ &&
+247801995 \eta ^5-119776230 \eta ^6+11764725 \eta ^7+848925 \eta ^8,
\\
Q_{30} &=& 2695+36450 \eta -392931 \eta ^2+909975 \eta ^3+3638115 \eta ^4
+3638115 \eta ^5+909975 \eta ^6
\nonumber \\ &&
-392931 \eta ^7+36450 \eta ^8
+2695 \eta ^9-8388608 \eta ^{9/2},
\\
Q_{31} &=& 2695+36450 \eta -392931 \eta ^2+909975 \eta ^3+3638115 \eta ^4+3638115 \eta ^5+909975 \eta ^6
\nonumber \\ &&
-392931 \eta ^7+36450 \eta ^8+2695 \eta ^9+8388608 \eta ^{9/2},
\\
Q_{32} &=& 121275+1680675 \eta -17110890 \eta ^2+35400285 \eta ^3-43091362 \eta ^4+35400285 \eta ^5
\nonumber \\ &&
-17110890 \eta ^6+1680675 \eta ^7+121275 \eta ^8,
\\
Q_{33} &=& -267411375-3646463625 \eta +38576020770 \eta ^2-86110332525 \eta ^3\nonumber \\ &&
+1036773146624 \eta ^4+86110332525 \eta ^5-38576020770 \eta ^6
\nonumber \\ &&
+3646463625 \eta ^7+267411375 \eta ^8,
\\
Q_{34} &=& -267411375-3646463625 \eta +38576020770 \eta ^2-86110332525 \eta ^3
\nonumber \\ &&
-1036773146624 \eta ^4+86110332525 \eta ^5-38576020770 \eta ^6
\nonumber \\ &&
+3646463625 \eta ^7+267411375 \eta ^8.
\end{eqnarray}
The above expressions depend on $\eta$ only, but not on $\sqrt{\eta}$, and they are symmetric under $\eta 
\leftrightarrow \eta^{-1}$. The expansions of the  $O(\varepsilon^0)$ 
terms for $N = 1,3,5$  up to $O(\eta^3 \ln^3(\eta))$ for $\eta < 1$ agree with the results we have found in
an independent calculation using {\tt Q2e} \cite{Harlander:1997zb,Seidensticker:1999bb}. 
With growing values of $N$, these expressions exhibit a growing degree of the polynomials in $\eta$. We have 
found that the general $N$ formula cannot be expressed as a sum--product solution by means of difference 
field theory. This means that the corresponding solution will be given by a higher transcendental function
depending on $N$ and $\eta$. Forming the corresponding numerical Mellin moments for the complete $x$-space expressions,
we agree with the analytic Mellin moments given in this appendix.

\newpage

\end{document}